\documentclass[a4paper,french,10pt]{article}
\usepackage{lineno}

\usepackage[T1]{fontenc}
\usepackage{amsmath}
\usepackage{array}
\usepackage{rotating}
\usepackage{textcomp}
\usepackage[francais,english]{babel}
\usepackage{enumitem} 
\usepackage{amssymb}  
\usepackage{graphicx}
\usepackage{epstopdf}
\usepackage{color}
\usepackage{epsf}  
\usepackage{url} 
\usepackage{tikz}
\usetikzlibrary{arrows,shapes,positioning,shadows,trees}
\usepackage{multicol} 
\title{Subrelativistic electron source for Dielectric Laser Acceleration, elements of design}
\author{Jean-Luc BABIGEON\footnote{Laboratoire de l'Acc\'el\'erateur Lin\'eaire - 91898 - Orsay, France; babigeon\at lal.in2p3.fr}} 
\date{} 
\begin{document}
\selectlanguage{english}
\maketitle

\tableofcontents
\newpage
\listoffigures 

\newpage

\section{Summary}

We describe and discuss solutions for an innovating sub-relativistic electron source, based on pulsed DC gun, including linear (1D) Field Emitter Array (FEA) associated immediately with Laser Dielectric Accelerator (DLA) stage. That setup is designed to avoid any standard accelerator component (electrostatic or magnetic macroscopic dipole \ldots) Parts of the diagnostics, including spectrometer and all diagnostics, are planned to rely on nano-structures. Then the entire setup -pulsed source, laser and measurements ends excepted- should be enclosed inside a decimeter range vacuum chamber. The goal of that paper is to suggest experimental orientations, which are discussed inside our laboratory.

However we shall also notice the theoretical issues among which we actually work, these one could be presented in a next work. 

\section{Dielectric Laser Acceleration for the future projects of high energy linacs}

Laser driven acceleration in Dielectric (DLA) principle is not a new concept, it has been described -at less- soon as 1996 for instance by \cite{Huang}. The \og{}Accelerator-in-chip\fg{}, under impulse of Pr Byer, has been schematically described by \cite{England}; assuming the occurence of thousand of low cost lasers and dielectric amplifiers, each amplifier of 1mm characteristic length, it could be possible, for example, to reach the TeV range, independently of numerous other industrial and research applications. This concept intends to solve following practical considerations or questions :

\begin{enumerate}
\item is it possible to stop inflating infrastructure sizes/costs for TeV electron accelerators ?
\item Can we exploit the emerging market of laser sources for establishing new concepts and performances in accelerators ?
\end{enumerate}

100TeV proton collider made by standard technology could amount up to 190km infrastructures with copper cavities, klystrons, standard magnetics... And added cumbersome cold technologies \cite{Richter}. The goal of alternative techniques is to reduced the infrastructures, at less for 1TeV electrons colliders, to much less than km range Linac.

In this contribution, we describe the very first stage 0-10keV, of a 1TeV Accelerator-in-chip, ie sub-relativistic electron source in the 10keV range. Indeed, it is considered to be the most difficult step, considering the spatio-temporal exploding character of electron bunch at low energies, particularly at the demanding total charge for that type of source. It is probably why first efforts were directed on relativistic dielectric stages, inserted inside conventional accelerators \cite{Mcneur}\cite{Plettner}\cite{Wootton}. From other side, if we demonstrate that in contrary, we are able to reach low emittance figures at low energy level, it will considerably help for instance, the implementation of low cost free electron lasers \cite{Ganther}.

Beyond emittance, the accelerator figure of merit is -considered to be- luminosity  so we are to combine emittance with enough charge. In a first introductory section \ref{principle}, we describe the principle of our femtosecond electron source, and specify the desirable performances, relying on field assisted photo-emission induced on Field Emitter Arrays (FEA).

In the following section \ref{components}, we analyze successively the components of our source, pulsed photo-cathode, photonics, dielectric accelerator, and finally measurements; we point and separate numerous technical and theoretical problems associated, considering that it should not be pertinent to develop specific theoretical point if the design is not prealably defined at less in their broad lines.

Finally in section \ref{conclusion}, we sketch a working schedule, and discuss about the ressources and efforts to devote to it.

\section{Analysis of available electron emission techniques}\label{principle}

    \subsection{Overview}

\begin{figure}[h]
\center
	\includegraphics[scale=0.6]{{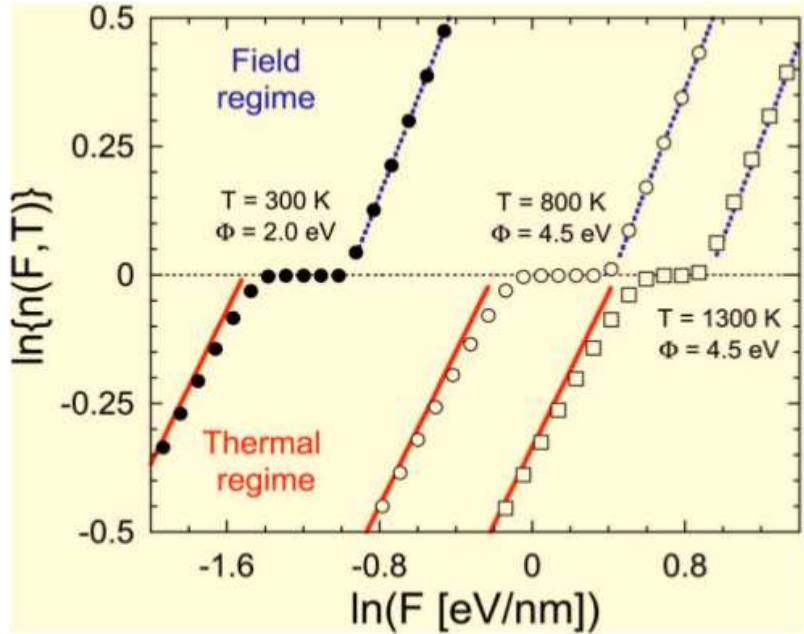}}
\caption{thermal and field emission regimes, and transition zones, after \cite{Jensen}}
\label{regimes}
\end{figure}

They are several possible regimes of emission in vacuum, cold field (tunneling), Schottky (thermally and field enhanced emission) and strong Field Emission (FE) (tunneling), depending on physical mechanisms. Instead plotting standard potential description like \cite{Dowell} (fig 2 of his lecture), we emphasize by figure \ref{regimes} from \cite{Jensen}, the intricate role of thermal component, which acts like a transition zone between pure Thermo-emission and Field emission. The n coefficient (see reference), $n=\frac{\beta_{T}}{\beta_{F}}$ is a marker of the predominance of each regime. $n >> 1$ is the field (F) emission domain, and $n << 1$ the thermal (kT) one. This interesting Jensen's idea may participate in the basis of FEA design, concurrently with \textit{abinitio} considerations.

In the general case, we extract the electron by electromagnetic potential difference, induced either by electrical voltage, or by laser photo-emission. In the following, we discuss the applicability of different regimes to our specific source.

    \subsection{Field emission induced by electrical voltages}

The strong FE regime is assumed to occur above a near field threshold in the range 3 to 10GV/m near FEA tips \cite{Bionta}. In that strong field mode, some \og{}thermal\fg{} effects occur.

The acceleration field in strong FE regime, is also prone to generate high charge, high energy anomalous electron in the front of the bunch,typically 20\% higher than normal arrival energy, in the case of Diode guns \cite{Levko}. With Diode guns, that outcome is interesting if we accept low repetition rates, with charges much larger than 1nC/bunch, given the time for the cathode to recover its original state, if it is yet possible, excluding electro-chemical induced effects. Although that type of source avoid the cost and use of a laser, it is rather oriented to high charge accelerators, like induction one, with nanosecond bunches. Some interesting experimental efforts -to follow if technology evolutes...- were devoted to production of sub-nanosecond electron pulses, with as high as several hundred of keV bunches \cite{Vyuga}. Anyway, the cathode recovering under such pulsed Ultra High Voltages seems to exclude more than 10Hz repetition rate, with limited cathode time life. We convinced ourself that it is also not trivial to generate picosecond emission with high PrF only by electrical pulses, so the choice is today oriented with laser assistance for ultra-fast bunches in accelerator -and spectroscopy- communities.

The inception of FE occurs in a tiny interval of voltages, due to non linear laws of emission. Under that threshold lies also Schottky mode. In that last mode, a few of electrons are emitted, either directly by the cathode or by the corrugations over the metallic walls of the gun, their distribution forming the \og{}black\fg{} current. Much electron sources for accelerators live with Schottky emission, RF standard guns included. However, the presence of FEA is expected to enhance cold FE against dark current. Moreover, regarding dielectric accelerators, the tiny dimensions of electron sources is one reason to choose a voltage range between 10 and 30kV partly because of breakdown consideration and/or thermal degradation. The voltage, in fact is obviously not the real marker, rather the emitted current. So it is admitted to use MV pulses, but with cm range cathode-anode distances, at the cost of low PrF. For instance, in \cite{Brau}, a distance apex-anode of 1cm for 50kV, with apex of $1 \mu m$ and tip length of 1cm, was linked with fields of $5 10^{9}V/m$. 

    \subsection{Field emission induced by laser}

Regarding photo-emission, many terms are also often used, field assisted, Schottky assisted, Photo-field photo-emission, above threshold ionisation, optical tunneling\ldots. The combination of electrical (HV) work and laser energy overcomes the work function barrier and permits electron emission in vacuum. 

Let's recall as in \cite{Bionta}, that the major criterion to distinguish strong FE induced by laser, from multi-photon like Schottky is the Keldysh factor.

To summarize, electronic emission may be generated \footnote{depending on amplitude, duration, source impedance, power \ldots of electrical or optical sources; for instance, with ultra short 1MV UHV electrical pulse, we may meet tunneling field emission type, while with 40kV $1\mu s$ electrical pulse with high current, Schottky type may occur, etc.}, by pulsed electrical source (diode guns), by HV source assisting photo-emission (DC guns or photo-field emission if electrical field is not present), or by photo-emission synchronized with electrical HV pulse (pulsed DC guns).

    \subsection{Our project : Field assisted photo-emission in pulsed configuration}\label{presentframe}

    \subsubsection{Principal features}
    
    That present contribution is oriented to field assisted photo-emission by a electrical pulsed-DC gun. The choice of pulsed voltage is suggested by the technological performances of high PrFs as precised above. \footnote{Some features are common with ultra-fast spectroscopy experiments} We have followed the same original principle than \cite{Ganter2} but with following modifications :

    \begin{enumerate}
    \item Laser is femtosecond instead of picosecond one,
    \item Nanosecond voltage is 10 kV ($\gg MHz$) instead at 60kV(30Hz),
    \item Voltage waveform is bipolar,
    \item Emitter is 1D CNT FEA instead of single ZrC tip, with individual apexes of 20 to 50 nm instead of $1 \mu m$, but length of the tips may reach $1\mu m$,
    \item There are two possible designs for the gun, coaxial one like \cite{Ganter2} and integrated with DLA
    \end{enumerate}

\footnote{in the following, we nevertheless shall denote the \og{}field emission\fg{}, the global effect of the two fields, emphasizing that total electric field sums laser induced \textbf{and} electromagnetic one, induced by pulsed electrical mode}

Despite its -at first glance- added complexity, the pulsed setup/mode offers several advantages :

\begin{enumerate}
\item the breakdown level is higher, as it is well explained by \cite{Wang}, who defines a specific criterion for transient AC fields, with levels beyond Kilpatrick threshold,
\item the combination of pulsed laser and (fast) pulsed voltage allows to isolate and cut the electron bunch to ultra short longitudinal emittances before the acceptance of DLA. Indeed, with the first DC guns setups used for demonstrator of DLA acceleration \cite{Mcneur}, it was shown that only a low fraction of the beam -emitted permanently in case of DC guns-, was in the proper phase for acceleration. While it was convenient for experimental evaluation, extrapolation to real source should ask for other solutions.
\item with a bipolar electrical pulse, we probably reduce the ionic bombardment on the cathode, because the mean value of accelerating forces on ions is null inside a bipolar cycle; furthermore, it could help to discharging the dielectric surfaces of the first accelerating stage, and so enhance the efficiency of acceleration,
\item one can easily show \cite{Dowell} that Schottky effect induced by pulsed source reduces the necessary gap for electrons to be emitted in vacuum  \footnote{but we admit with that author, that we cannot reduce energy gap to zero!}; either one can see it as improvement of quantum yield of photo-emission, for a given laser wavelength, either as a technique for working with the same quantum yield, but in the near infrared range, instead of costly near UV range, 
\item fast pulsed cathode mode is anticipated to lighten other physical mechanisms that electrostatic field of the DC gun; we shall come back on it, in section \ref{components}, precisely inside design of photo-cathode. 
\end{enumerate}

\begin{figure}
    \begin{minipage}[t]{12cm} 
        \centering
        \includegraphics[width=11cm]{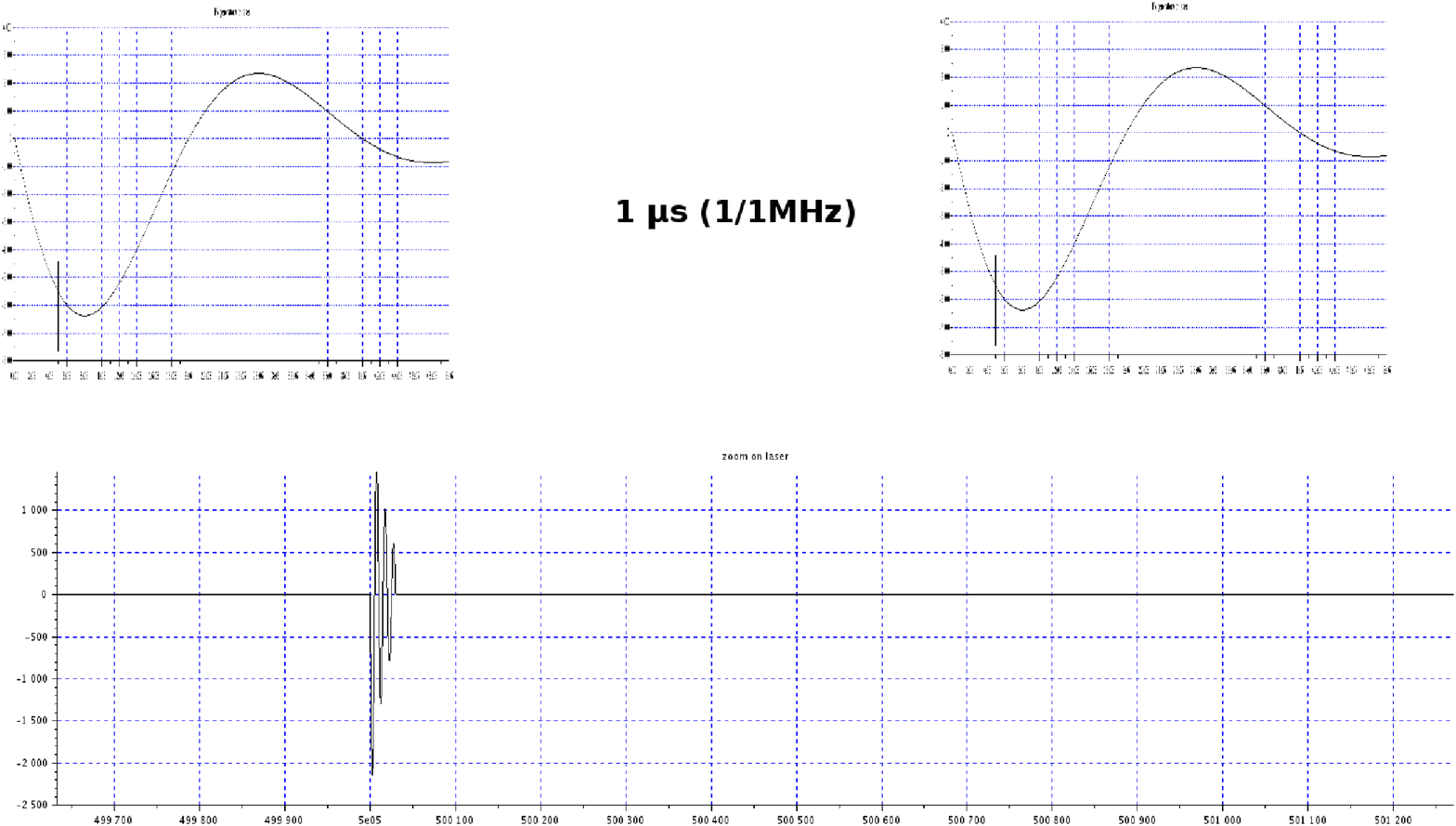}
        \caption{Optical pulse combined with HV pulse}
        \label{HVsum}
    \end{minipage}
    \begin{minipage}[t]{12cm}
        \centering
        \includegraphics[width=11cm]{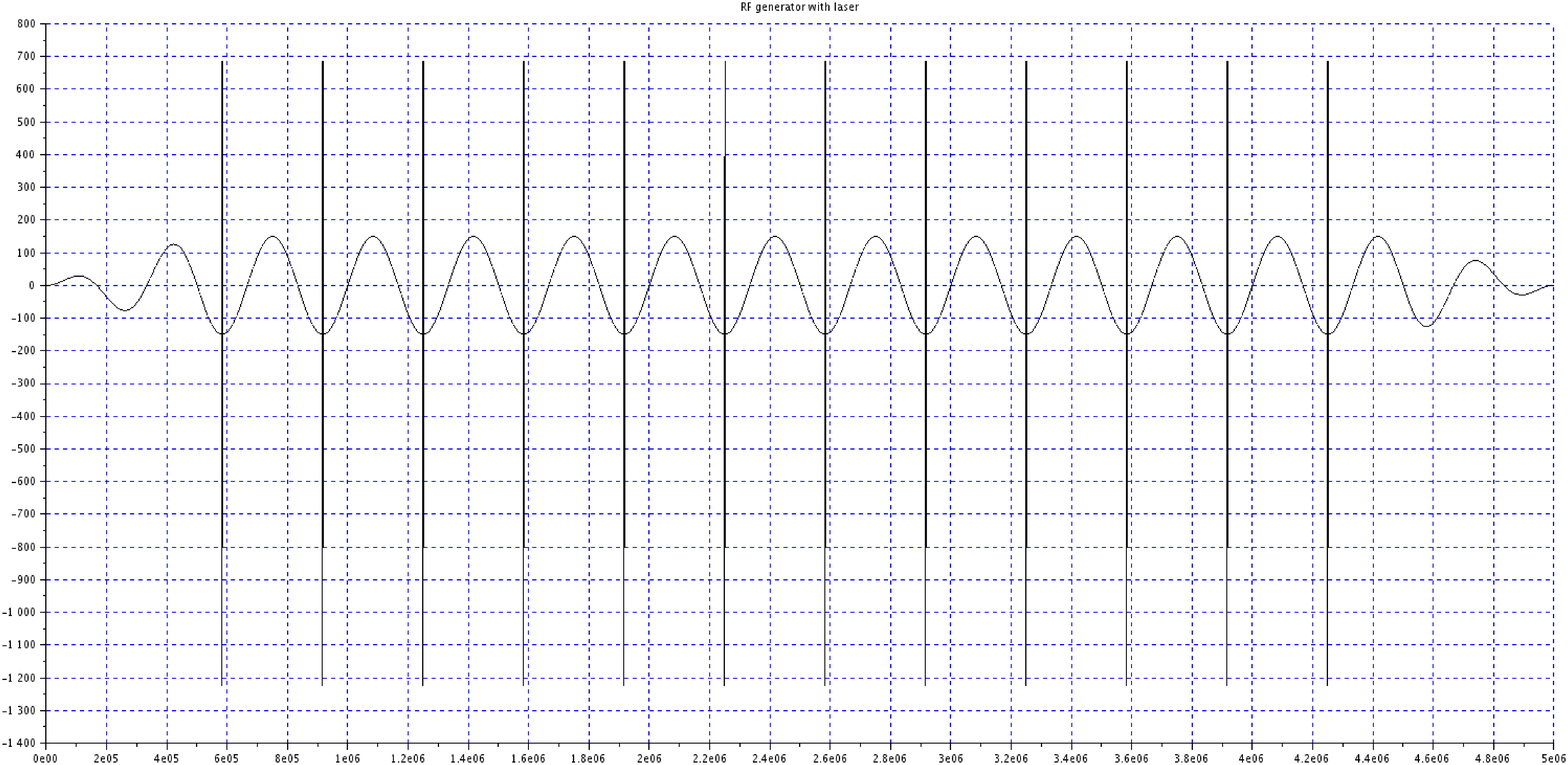}
        \caption{Optical pulse combined with RF pulsed amplifier}
        \label{RFsum}
    \end{minipage}
\end{figure}

The superposition of electrical and optical pulses is shown in figure \ref{HVsum} and \ref{RFsum} in the two cases of HV pulsed and RF pulsed sources. Of course, technology may enlarge the limits, given by thermal and power considerations. We represent the expected waveform diffracted by FEA, not the initial laser Waveform, which could be for instance a Gaussian one. Indeed, it is very probable that any unipolar waveform is not conserved by interaction with high pass elements like monopoles/dipoles of the FEA. Note that these representations are for demonstrative goal only. In HV figures, range is ns, and in RF one, $\mu s$.

In our study of DSRD pulsed generator, we had not precised the exact width of the \og{}flat top\fg{}. In fact, with a single stage, we expect a width of 10ns, with transition times 0.5ns approximately. With a second compression stage, we estimate the waveform to be finally 10kV/1ns/200ps, these figures explaining by themselves. So in the figure hereunder, we took in hypothesis that last performance. The PrF of the HV generator is taken to be 1MHz. Laser oscillator available at Lal has 80MHz PrF, ie 13ns inter-pulse period, so we have to divert 1/80 of laser impulses.

For the RF pulsed amplifier, we have borrowed the model of 400W S-band amplifier for klystron, with 2 microsecond width. In that case, we can inject roughly 10 laser pulses inside one RF pulse.  Of course, these figures are not to be definitive, for instance, an X-band amplifier is conceivable and power is chosen by availability criterium. Our RF amplifier design capabilities are given by the transistor one, ie $100 \mu s/400W/20\%$, which result in 2kHz PrF. We notice that only $2\mu s$ were specified, not 100, the reason why being our care about possible thermal effect inside amplifier. If we keep anyway, the PrF, we can extrapolate the performances of each present power sources, regarding the number of pulse/s, and the Lal capabilities.

\begin{enumerate}
\item in HV case : 1Mhz with 1/80 laser pulses, in multi-burst mode, 20 pulses by burst with 100Hz; in the hypothesis on dividing laser frequency, we may expect 20x100 = 2000 pulses/second,
\item in RF case : 10 laser pulses inside 1 RF one, Prf 2kHz, so 10x2000 = 20000 pulses/second.
\end{enumerate}

\textbf{*} : One may ask if space charge, considered as minor inside a single pulse, could not occur with high PrFs. But the principal issue today felt is the possibility of induced electromagnetic effects from inter pulses or bunches-structures interactions.

\begin{figure}[h]
\center
	\includegraphics[scale=0.4]{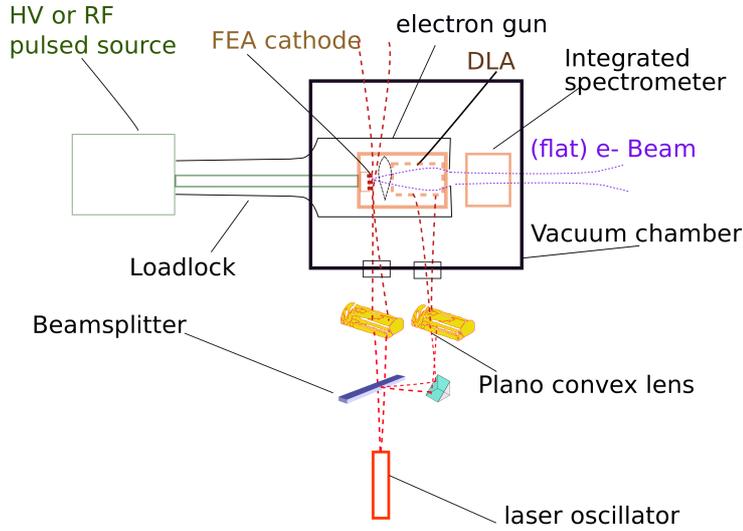}
\caption{DC-pulsed principle}
\label{DC_pulsed}
\end{figure}

The figure \ref{DC_pulsed} describe schematically the pulsed-DC gun technique.

    \subsubsection{Consequences on the phase choice of laser}

    As the laser pulse is superimposed on electrical one, guessed perfect synchronism, it is known that efficient temporal compression of the bunch occurs depending of relative phase of laser injection. The Forbes non linearity of Schottky emission could help that compression. However, in HV case, it becomes then impossible to work with multi-pulse concept, placing several laser pulses on \og{}flat top\fg{} of nanosecond electrical pulse, because evidently all pulses become different.

    Moreover, there is a supplementary possibility to explore with tuning the phase, ie to study the Schottky effect occurrence linked with 1 or multi photon extraction, and very low emittance \cite{Ysof}, knowing we opt for a non energetic laser regime, where thermal effects are to be minimized and Keldish factor high. All these considerations are known in electron source domain. However they are to be faced to new experimental design.

\section{Components of sub-relativistic electron source}\label{components}
    
    \subsection{Pulsed electrical source}

\begin{figure}
    \begin{minipage}[t]{12cm}
        \centering
        \includegraphics[width=11cm]{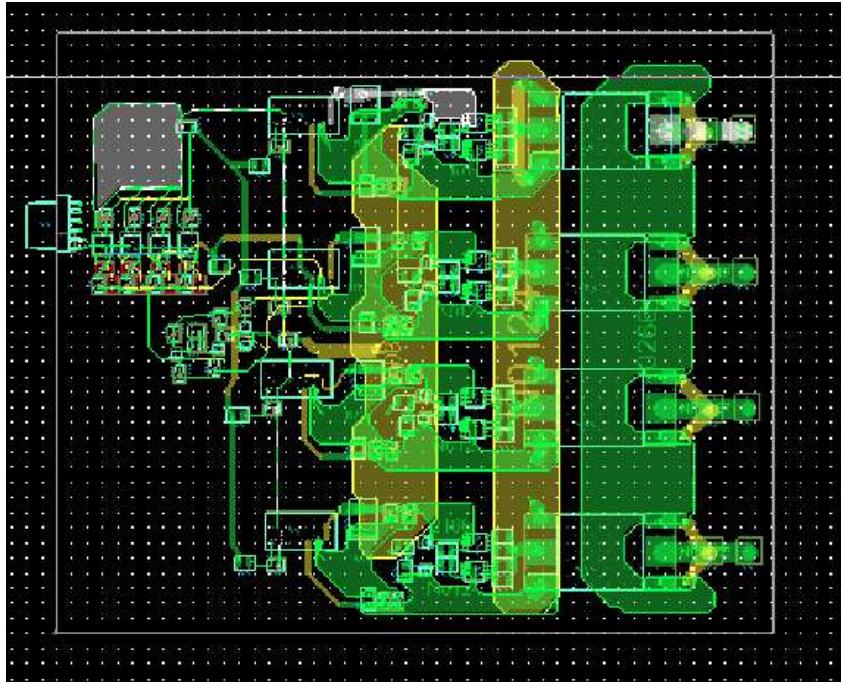}
        \caption{Layout of primary current switch}
        \label{routage}
    \end{minipage}
\end{figure}

Pulsed source is a  High Power Pulsed (HPP) High Voltage (HV) generator, with sub nanosecond rise and fall times. In its principle, it follows the DSRD (Drift Step Recovery Diode) generator made by \cite{Akre}. We have studied such a generator at Lal \cite{babigeon}, the objectives being 100ps rise time at PrF higher than 1 MHz and for peak voltages of 10kV (which means peak to peak voltage of 20kV). The primary current PCB layout of the generator with 4 Mosfet switches in parallel is represented on figure \ref{routage}.

\begin{figure}[h]
\center
	\includegraphics[scale=0.4]{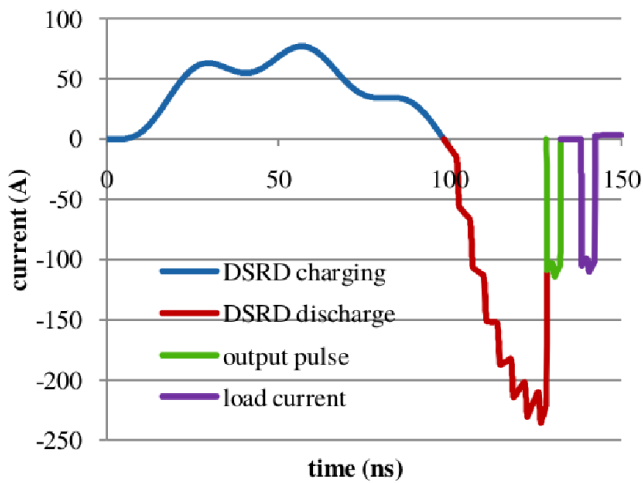}
\caption{Direct current and pulse timing of DSRD generator, from \cite{Akre}}
\label{timing}
\end{figure}

We have also analyzed the reduction of the jitter. With conventional fast electronic techniques, it seems possible to reach a $< 5ps$ jitter. A lower one could be achieved, for instance in subps range, at the expense of rigorous voltage regulation, which is explained below. Indeed, the control of direct pulse current is fundamental for the DSRD diode. We show at figure \ref{timing}, thanks to \cite{Akre}, the timing of pulse generation.

\textbf{*} : Let's come back to timing diagram of the pulse, in the figure \ref{timing}. We can see qualitatively that the jitter will depend of the position of Primary current. In fact, if the current waveform is perfectly reproducible, the precise instant of current reversal is determined, and the diode response being somewhat deterministic compared to times scales, we can admit that jitter will be minimal, of course implicitly accepting a precise trigger. So the timing question depends principally of \textbf{precision of current injection}, and specifically of precision of maximum level, or maximum level of voltage charge.

The emergence of new SiC HV diodes \cite{Huang2} brought us to investigate the semi-conductor physics to determinate if their use in DSRD generator is possible. Indeed, SiC diodes are presented like soft recovery rectifiers, for power market; it is the opposite quality we need, ie a component able to store energy from Direct current and give it back in a kind of \og{}compression mode\fg{} during the opening phase. That concept of opening switch associated with inductive generator is a motivating technology. The dynamic behavior of the switch depends mainly of metallurgical profile of doping. A lot of convincing hypotheses bring us today to estimate with Grekhov arguments \cite{Grekhov} that the P doped PIN diodes, P+pN are better fitted than P+nN one for DSRD effect. It has to be scientifically ascertained by futures simulations, physics models and specific SiC prototypes.

Nevertheless, an other issue is the repetition frequency. Despite its tremendous value of 1MHz for DSRD HV generators, higher than any PrF of conventional capacitor discharge HPP HV pulse generators  the requirement of accelerator community for luminosity could ask for 100MHz to GHz range of PrF (see section \ref{presentframe}; it is a serious challenge.

Some recent analysis drives us to consider the use of power RF amplifier. Let's, in first, look at necessary power : in case of $50\Omega$ adaptation, the instantaneous necessary power is $P=\frac{\frac{V^{2}}{2}}{Z}= \frac{10kV^{2}}{4*50}= 500kW$. Of course, in case of CW amplifier, instantaneous power is of same order than mean power. If we wish to build it in L or Sband for instance, it is far from economical today possibilities, and not coherent with \og{}accelerator-in-chip\fg{} philosophy. Note that the advantage is here to pulse generator, considering instantaneous power.

Nevertheless, two factors may reduce the minimum power :

\begin{enumerate}
\item the laser PrF are rather in 100MHz range for the moment,
\item the input impedance of our gun is much higher than $50\Omega$
\end{enumerate}

Independently from collider specifications, the consequences of PrF reduction is that power will scale with PrF, so for typical 80MHz laser oscillator (Element 100 Ti:Sa from Spectra-Physics), the ratio is $\frac{P_{80MHz}}{P_{3HGz}}=\frac{80MHz}{3000MHz}= 0.266$ ie the necessary (here mean) power will be 133kW, which remains very high. We have now to evaluate the real input impedance of our cathode.

    \subsection{Adaptation of input cathode impedance}

\begin{figure}
    \begin{minipage}[t]{15cm}
        \centering
        \includegraphics[width=14cm]{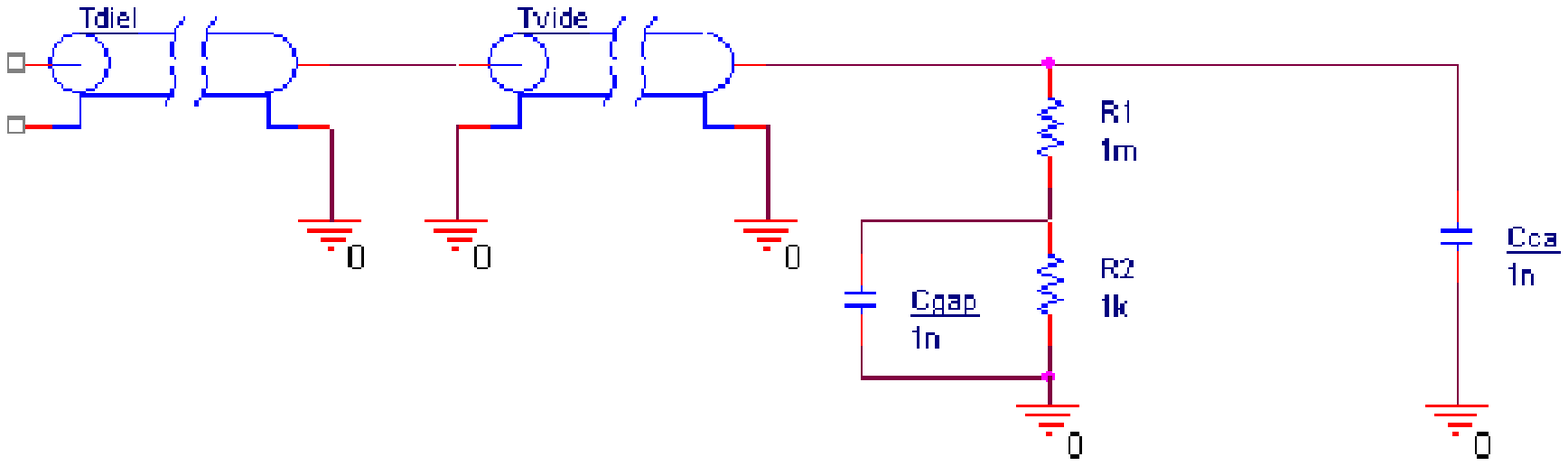}
        \caption{Equivalent circuit of the coaxial gun}
        \label{diode_equivalent}
    \end{minipage}
    \begin{minipage}[t]{15cm}
        \centering
        \includegraphics[width=14cm]{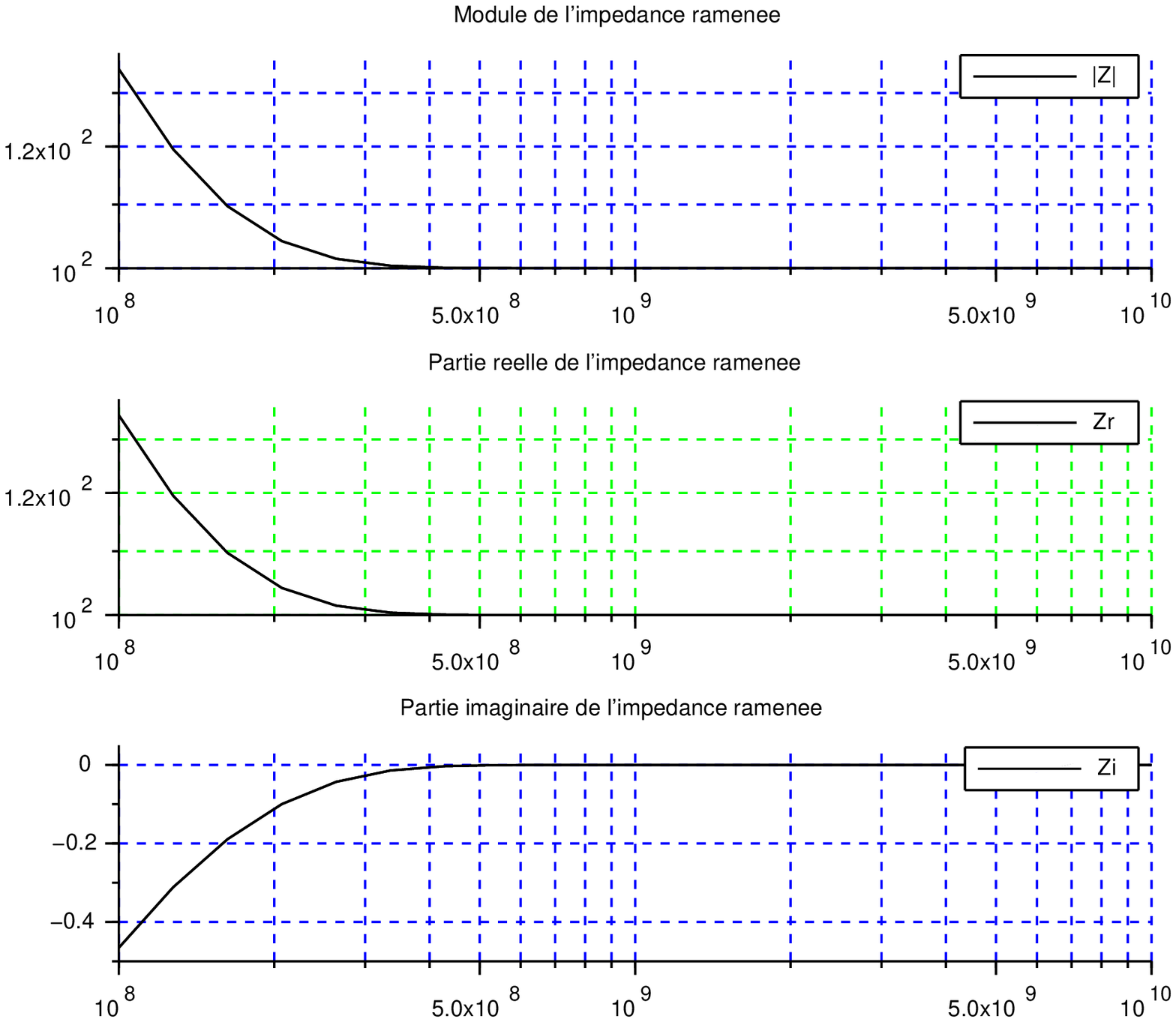}
        \caption{Input impedance of the coaxial gun}
        \label{zin_canon}
    \end{minipage}
\end{figure}

  The geometrical and electrical models of our field emission cathode are represented on figures \ref{diode_equivalent} and \ref{zin_canon}

  Here $C_{gap}$ is cathode-anode capacity, assimilating it to a plate condenser, ie $C_{gap} \sim 8.85 10^{-2}pF$ with the typical dimensions of our gun. If we see roughly an emission tip as a cylinder, the DC resistance is $R_{e} \approx 0.35 m\Omega$. We calculate also the Forbes non linear resistance for vacuum emission of that tip. \footnote{In vacuum,the corresponding wavelength to 100ps rise times is 3cm. It is the characteristic dimension of our electron gun. So, using the quasi-static approximation we are to stay aware of that}. With that circuit model and the associated equivalent input admittance we computed the time response of current density, under application of electrical bipolar pulse. We represent in the figure \ref{zin_canon}, the equivalent input impedance, including the short vacuum transmission line inside our loadlock, and an arbitrary length of transmission line between generator and loadlock.

  The figure \ref{zin_canon} represents real and imaginary parts of that input impedance, viewed from the entrance of the coaxial injection line, centered around 300MHz frequency of the signal spectrum, in the case of HV pulsed generator.
  
Note that as wavelength is 3cm, it is expected that input impedance should pass by minima and maxima. The idea is to stay at a high level of returned impedance, say $1k\Omega$ for instance. The advantage is that generator current is minimized. In other side, the allowed current to the tip is to be regulated, or limited by the source, in order to avoid thermal runaway and to improve reproducibility of emission. Of course, the precise impedance is not specified here, because the final design includes many tip emitters \footnote{inside our approximate study, it is not so difficult to incorporate N emitters in parallel, admitting the mutual distance between them sufficient to neglect mutual impedances}, but it is a qualitative and demonstrative point.

An other conclusion of our approach, is that whatever the injection mode, by pulse HV generator or RF pulsed amplifier, the adaptation condition is mandatory, and it implies a injection length line well defined, at less cm precision. Of course, it is always possible to add a last tuning element, for instance a cage capacitor. Last but not least, that adaptation should be better optimized.

\textbf{*} : Adaptation is not of course a single figure, because at less, the Forbes impedance is level dependent during the bipolar pulse. Then, the determination of precise adaptation is not a trivial question, but a synthesis problem in linear or non linear scope. As it is well known, even the linear solution is not univoque, depending on transfer function choice. Among candidates, could be elliptical Jacobi transfer function because they assume the steepest frequency roll-off, so dispersion inside crystal is minimum.

With $1k\Omega$ impedance instead $50\Omega$, the ratio being 20, the necessary power of our source, evaluated at precedent section,  could be $\frac{133kW}{20}=6.65 kW$. With 20\% rate of the signal, we only need 1.33kW.

 To summarize, Pulsed HPP HV generator versus RF power amplifier presents some -known- advantages and drawbacks

  \begin{enumerate}
  \item For medium and high instantaneous powers above, say some kW, pulsed HV, or High Power Pulsed generator have undisputed yields, probably 90\% against 60\% for standard RF class, and they can reach 100kV range and more,\cite{Hendriks},
  \item For PrF above 100MHz, difficulties will arise for implementations of FPP generators,
  \item synchronization and jitter remains a common issue for both solutions,
  \end{enumerate}

  \subsection{Nano-structured photo-cathodes and Field Emitter Arrays (FEA)}

Since the first nano-structured cathodes by Spindt, much pioneering work has been undertaken for instance by Dr Tsujino and its team for now twenty years \cite{Tsujino}. They tested the diode and triode configurations. In triode one, a gate allows us to trigger a fast pulse, avoiding laser photo-emission. The performance of such triodes are today limited to picosecond bunches. The diode configuration, where the bunch is effectively emitted by femtosecond laser, has been also studied by Toulouse with a DC gun \cite{Bionta}-generally made with a microscope 100kV basis- and by Hommelhoff team. In that section, we focus on cathode technology, ie the materials and structure.

\begin{figure}
    \begin{minipage}[t]{7cm}
        \centering
        \includegraphics[width=6cm]{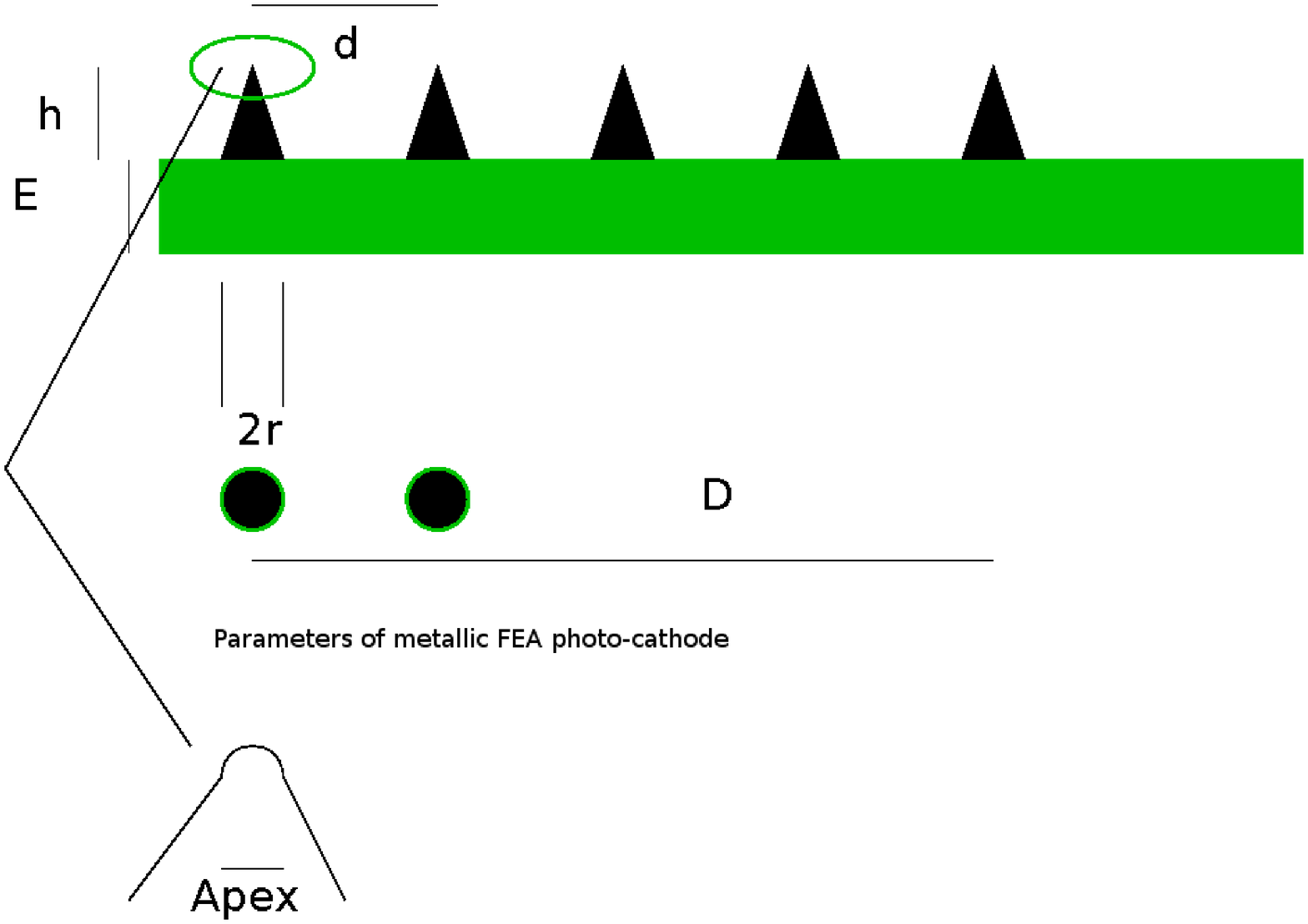}
        \caption{Spindt-like FEA 1D Cathode}
        \label{spindt}
    \end{minipage}
    \begin{minipage}[t]{7cm}
        \centering
        \includegraphics[width=6cm]{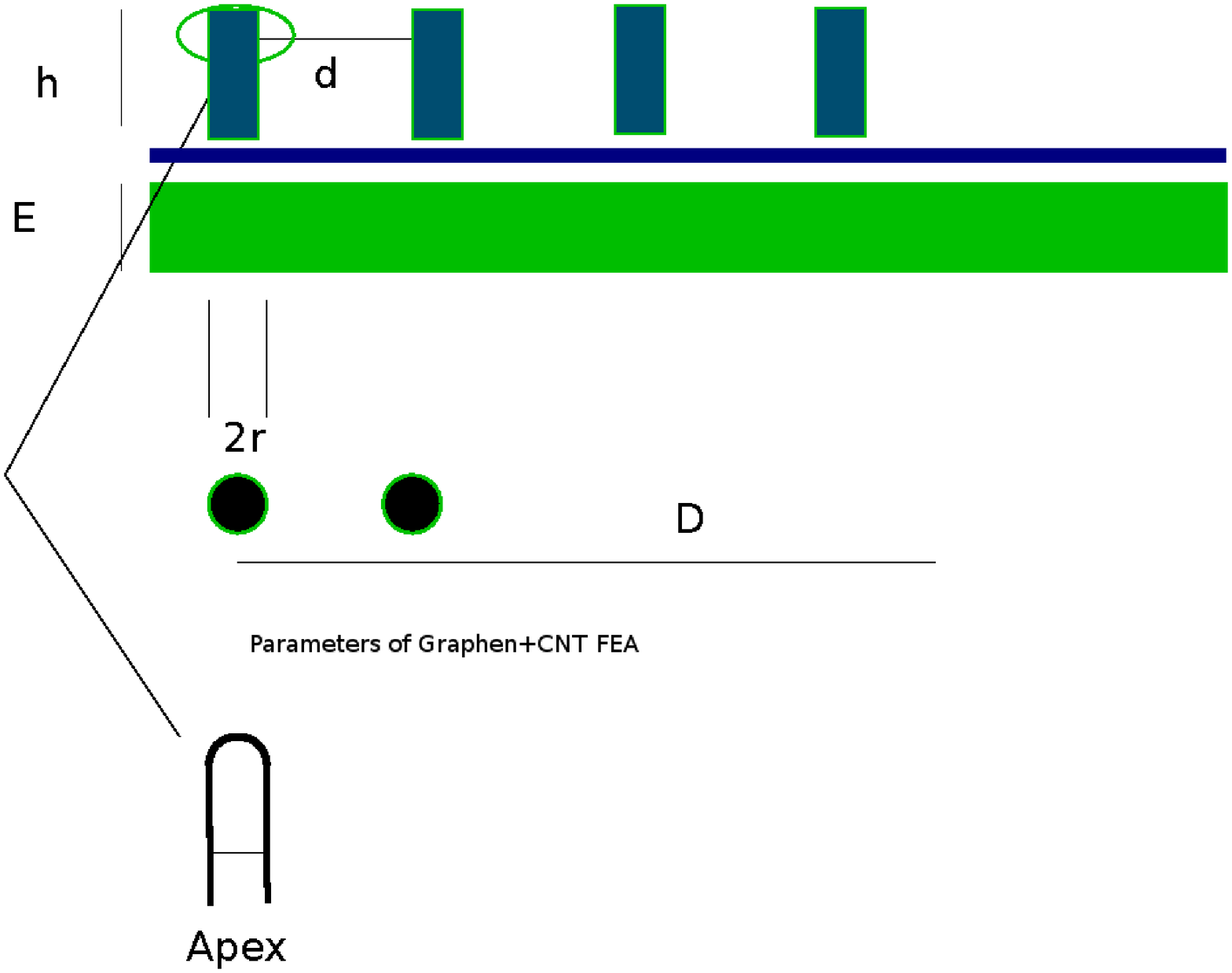}
        \caption{Carbon Nano-tube FEA 1D cathode}
        \label{cnt}
    \end{minipage}
\end{figure}

Two configurations are showed in figures \ref{spindt} and \ref{cnt} : a spindt-like one, and a new proposed one, relying on Carbon family. Spindt-like cathode is studied far from now  \cite{Spindt}, \cite{Kirk} and constitutes a good reference comparison to any new candidate.   

\textbf{*} : Since we have insinuated that some additional physics must be considered in sub nanosecond pulsed-DC guns, we shall only mention here, non thermal/cooling Nottingham effect\cite{Keser} and plasmonic field enhancement, at the tips level. However, \textbf{before} describing the spatial domain around the tips, at the immediate vicinity of vacuum, it is fundamental to go \og{}inside\fg{} the -photo-cathode and describe its physics with help of \textit{abinitio} simulations. Indeed, phonons and surface state contributions for instance, strongly modify moment and energy distribution at that immediate interface.   

The new photo-cathode, based on carbon family, differs from initial Tungstene, Molybdene, \ldots Spindt FEA by many aspects :

\begin{enumerate}
\item a thin layer of graphene in Bernal configuration,
\item growing on graphene, of regularly spaced CNT in a 1D (linear) FEA
\end{enumerate}

\textbf{*} : as it is intuitive that a 1D linear array will perform optimally in emittance, in one direction X and badly in Y, the quantitative gain has to be better explained.
    
    It is expected that graphene constitute an electron reservoir for the CNT. The transition between conductive cathode and CNT needs to have some electrical resistance, in order to limit current in CNT. Incidentally, graphene layer protects against eventual back ionic collision on the cathode. CNT are proposed because of their outstanding thermal conductivity, and their ratio height:diameter, raising more than 1000 \cite{Li}

    Several questions arises against that proposal, among them :

    \begin{enumerate}
    \item considering copper substrate coming from our gun, how will be the compatibility between copper and graphene (see some answer elements in \cite{Santanu}) ?
    \item how to grow a regular array of CNT on graphene (see some answer elements in \cite{Niu}) ?
    \item how are the characteristics of graphene and CNT for efficient emission ?
    \item can we ascertain these characteristics in fabrication process ?
    \end{enumerate}

    \textbf{*} : although these questions may be appreciated mainly as fabrication specific, they have their counterpoint in theoretical and computational sides. Indeed, for instance, the compatibility copper-graphene is a wave-function problem, and is studied with \textit{abinitio} tools \cite{Gonze}. The CNT performances also, ie thermal capacity and 1D transport features are also belonging to solid state physics and described by the same tools \cite{Hiebel}.

    \textbf{*} : One issue is the representation of ultrafast transients with the wave function using TDDFT \cite{Botti}. Last but not least, excited states and there hamiltonian representation -if correct-, are a theoretical today topic.
    
Let's describe qualitatively the expected behavior of the assembling copper-graphene-CNT under high electric fields. In static simple description, there is a voltage induced between each nanotip and the graphene layer in its immediate proximity. It is furthermore guessed that a voltage will be present between each nanotip and copper substrate. So the graphene layer is prone to be stressed by electrostatic field. In that situation, some induced \og{}doping\fg{} appears \footnote{a description of induced doping in such a Dirac crystal like graphene is of course very rough. The correct treatment could be the description of Brillouin zones, and the derivation of normal modes}. These doping zones are to be localized only at the very near bases of CNT and are similar to pn junctions, n side being the substrate, and p side the base of CNT.   

The CNT growing on several interface is a recent topic. However, some partial results seems to show a good adhesion with graphene layer. The graphene in itself may present a very different mechanical resistance, its behavior depending on his structure and layer numbers.
 
The first expected results of such a combination, could be the following :

\begin{enumerate}
\item thermal coefficient being 10 times those of copper, the capability to raise the melting temperature over the critical one, so Nottingham effect becomes a cooling mechanism,
\item form factor being extremely big, will be associated with a great $\beta$ factor, so the effective electric field will be huge, with a low injected power,
\item electrical resistance of CNT is also expected to be much lower than metals, and a resistive graphene layer is desirable, in order to limit saturating currents in emitters,
\end{enumerate}

    \subsection{nano displacements and spatial alignments between components}

    Several components need to be aligned with a great precision :
    \begin{enumerate}
    \item FEA cathode and optics -if any-,
    \item FEA cathode and/or optics with DLA,
    \item Laser beam with FEA,
    \item Laser beam with DLA
    \end{enumerate}

An interesting setup was described by \cite{Quinonez} in their point projection microscope, they decided to nano-position only the laser optics. We notice however that they have not ultra precision criteria between the emitting tip and the specimen.  
    
    Before describing our proposal, which is finally similar to that reference, let's explore the possibilities of nano-positioning DLA inside chamber, in link with the best industrial performances. We have analyzed the simple case of aligning the FEA cathode with a standard DLA, for instance a double grating \cite{Breuer}. We have taken for DLA, a standard option of double grating, but any new idea should probably result in similar conclusions.

\begin{figure}[h]
\center
	\includegraphics[scale=0.4]{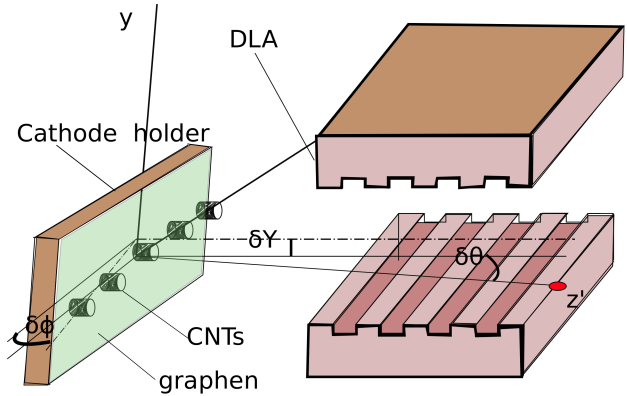}
\caption{geometry and freedom degrees in case of independent components inside gun}
\label{nano_pos}
\end{figure}

    Available precision motions, widely used in microscopy, are piezo electrical translators and rotators, and there are possibilities to assemble them by modularity. Let's consider the precedent simple scheme in the figure \ref{nano_pos}. 

If we want to align properly that set, we must manage a priori these degrees of freedom :lateral X, lateral Y, longitudinal Z displacements relatively to the front entrance of the DLA, and the 2 angles $\theta$ and $\phi$ of the beam relatively to the DLA axis. In fact we remind that we work with flat beam, so displacement X is not so much precision demanding, we can forget it. We can also work with a fixed $Z=Z_{0}$ so forget also in Z freedom. Finally, there are 3 freedom degrees left, Y, $\theta$ and $\phi$. Accounting Breuer analysis and practical fabrications, the DLA entrance is supposed to be some hundred of nanometers, and -if we play with evanescent waves, the precision of localization is approximately 10nm, in a tuning range of 50 to 150nm, then it is the specification for Y.

The typical DLA length being $L_{DLA}=25\mu m$ in \cite{Breuer}, the first maximal deflection angle $\theta$ is given by $tg \frac{\theta}{L_{DLA}} < \frac {20nm}{25\mu m} \sim 10^{-3}$, so $\theta_{max}=1 mrad \sim \frac{1}{20} degree$, which is announced as inside performance of rotating piezo components.

The sensibility from other angle $\phi$ must be investigated further because it has an action on the 3 components of electromagnetic force impinging on electron and may be responsible of shifting phase and lateral acceleration.

Let's see again the Breuer's work with his notations, equating phase velocity of wave and particle leads (page 19 of \cite{Breuer}) to :

\begin{equation}\textbf{v}_{ph}.\frac{\textbf{v}}{\textit{v}}=\frac{\omega}{k_{||}}cos\phi = \beta c\end{equation}

    so \begin{equation}k_{||}=\frac{\omega cos\phi}{\beta c}= \frac{k_{0}cos\phi}{\beta}\end{equation}

    It is slightly different from Breuer result, because of presence of $cos\phi$ on numerator, not denominator. The results are, to that step, identical on only when $\phi=0$. We go on by defining

    \begin{equation}k_{\perp}=k_{0}\sqrt(1 - \frac{cos^{2}\phi}{\beta^{2}})= k_{0}\sqrt(1 - \frac{cos^{4}\phi}{\tilde{\beta}^{2}})=\frac{k_{0}}{\tilde{\beta}\tilde{\gamma}}*\sqrt(1 - \tilde{\gamma}^{2}(1 + cos^{4}\phi))\end{equation}

    where $\tilde{\beta} = \beta cos\phi$ and $\tilde{\gamma} = (1 - \tilde{\beta}^{2})^{-\frac{1}{2}}$

    We stop here and see what it implies on physical insight. We have $\tilde{\beta} <1$ and $\tilde{\gamma}$ is real, so it is yet pertinent to speak of evanescent waves and define a distance $\delta$ such we find for $\phi=0$. We also note that for $\phi \sim \frac{\pi}{2}$, the parallel wave component becomes negligible, and the wave tends to evolve to plane wave, with dispersion $k_{\perp}^{2} - \frac{\omega^{2}}{c^{2}} \sim k_{0}^{2} - \frac{\omega^{2}}{c^{2}} = 0$, the consequence being that acceleration is \textit{a priori} not possible in that configuration. \footnote{we can also make another change of variable, defining $\tilde{\beta}=\frac{cos\phi}{\beta}$. The dispersion relation $k_{\perp}^{2}=k_{0}^{2} - k_{||}^{2}$ drives us to $k_{\perp}=\frac{k_{0}}{\tilde{\gamma}}$, with same algebraic manipulations. However, here $\gamma \in [-\inf,+\inf]$, so existence of direct and evanescent waves is also shown possible}

    \begin{equation}\delta \sim \frac{i}{k_{\perp}} = \tilde{\beta}\tilde{\gamma}\frac{\lambda}{2\pi}\end{equation} 

    Here we still globally agree with the referenced results, defining evanescent regime with the same expression for the distance $\delta$, but only for $\phi=0$, which is of practical importance.

    But there are some divergences when we want to compute the electromagnetic fields, specially its sensibility to angle $\delta\phi$, with $\phi \neq 0$ in absence of particle. There are several directions to estimate them.

    \paragraph{geometrical analysis}
    
    In that analysis, we follow Breuer's analysis, and with a similar computation, we obtain :

    \begin{multicols}{2}\noindent
    \begin{equation}
      \textbf{E}=
      \left(
      \begin{array}{c}
        \frac{c}{\tilde{\beta}\tilde{\gamma}}\sqrt(1-\tilde{\gamma}^{2}(1+cos^{4}\phi))*B_{y} \\
        E_{y} \\
        -\frac{c*cos\phi}{\tilde{\beta}}*B_{y}
      \end{array}
      \right)
    \end{equation}
    \begin{equation}
      \textbf{B}=
      \left(
      \begin{array}{c}
        -\frac{1}{c*\tilde{\beta}\tilde{\gamma}}\sqrt(1-\tilde{\gamma}^{2}(1+cos^{4}\phi))*E_{y} \\
        B_{y} \\
        +\frac{cos\phi}{c*\tilde{\beta}}*E_{y}
      \end{array}
      \right)
    \end{equation}
    \end{multicols}

    Next step is to compute the force \begin{equation}\textbf{F}=\textbf{E}+\textbf{v}\times\textbf{B}\end{equation} and obtain

       \begin{equation}
      \frac{\textbf{F}}{q}=
      \left(
      \begin{array}{c}
        c*\alpha_{z}*B_{y} + sin\phi*E_{y} \\
        E_{y}(1 - cos\phi) \\
        -\frac{c^{2}}{v}*B_{y} + E_{y}*(1 + \beta sin\phi)
      \end{array}
      \right)
       \end{equation}

       where we have taken $k_{\perp}=(\alpha_{x},\alpha_{y},\alpha_{z})=(\frac{cos\phi}{\tilde{\beta}},0,\frac{1}{\tilde{\beta}\tilde{\gamma}}*\sqrt(1 - \gamma^{2}(1+cos^{4}\phi)))$

       and $\alpha_{z}$ is purely imaginary number.

       That result, expressed in wave coordinates -same conclusion applies in particle coordinate- seems to show that :

       \begin{enumerate}
       \item although magnetic components are out of phase for x and z components of the force, electric components are not strictly always out of phase in that case,
       \item there is a transverse force associated to accelerating one
       \end{enumerate}

       Letting out of discussion the vertical force, which could be balanced in a double grating, we have to evaluate the ratio of $\frac{F_{y}}{F_{x}}$ in the particle coordinates.

       Projecting it, and developing $\alpha_{z}$ in powers of $\phi$ for low incidence angles, we find that ratio :

       \begin{equation}\alpha_{z} \sim i*\frac{sgn(\phi)}{\beta}\sqrt(1 + \beta^{2})*(1 - \frac{1}{2}*(1+\beta^{2})*(\phi^{2}+\ldots))\end{equation} and finally

  \begin{equation}\frac{F_{y}}{F_{x}} \sim tg\phi*\frac{E_{y}(1 - cos\phi)}{c}*\frac{1}{ i*\frac{sgn(\phi)}{\beta}*\sqrt(1+\beta^{2})*(1 - \frac{1}{2(1+\beta^{2})}*(\phi^{2}+\ldots)*B_{y} +sin\phi*E_{y}}\end{equation} then finally

  \begin{equation}\frac{F_{y}}{F_{x}} \sim \phi^{3}*-i*sgn(\phi)*\frac{E_{y}}{c*B_{y}}*\frac{\beta}{\sqrt(1+\beta^{2}}\end{equation}

  In a nano-positionning scheme inside the cell, we see then that the deflecting force is out of phase with accelerating one -what was already guessed - but of high order with $\phi$ for low deflection angles. It is a positive result which shows that stability and precision may be reach, but in other side we asked dimensioning to a supplier, and minimum dimensions of an assembling of precision translators/rotators was announced to be at less 10cm, which represent unfortunately the major part of our chamber.

It is why we propose a complete integration of photonics -FEA and DLA- by an overall fabrication, and we report the freedom degrees on laser lenses, so there will be a complimentary evaluation to do with that scheme.
  
\paragraph{Some glance for more adapted analysis}
    
    Inside that approach, we estimate that the (double) grating configuration are subject to the scattering theory with laser wavelength $\succeq 1\mu m$ greater than grating characteristic dimensions $\lambda_{p} \succeq 200nm$, in the near field zone, but in neither of the classical near field zones, static, induction one. \footnote{Note also that the femtosecond laser pulse and bunch don't behave like stationary regimes, and wave packet formalism seems necessary. However, we shall stay inside frequency decomposition, assuming an adiabatic system evolution.}

    Let's consider the same geometry as figure \ref{nano_pos}

The coordinate center is taken in the center of a grating;

The vector potential may be written as \cite{Jackson} :

\begin{equation}\textbf{A}(\textbf{x})=\mu_{0} ik \sum_{lm} h_{l}^{1}(kr)Y_{lm}(\theta, \phi)*\int J(\textbf{x'})j_{l}(kr')Y_{lm}^{*}(\theta', \phi')d^{3}\textbf{x'}\end{equation}

where $h_{l}^{1}, j_{l}, Y_{lm}$ are respectively spherical Bessel of third kind, spherical Bessel of first kind and spherical harmonics functions \cite{Abramowitz}, $J(\textbf{x'})$ the elementary current of source \textbf{x'}, observed at \textbf{x}. The ranges of kr is $kr > \epsilon$ where $\epsilon$ is physically a minimum distance between the beam and the edges of grating.

We can already guess that the forces acting on particle are growing at the DLA entrance, to reach a maximum at the center, and decreases at the output. In fact, there are several domains of approximations depending on quantities kr and kr'. At the extremities, $kr \leq 1$ and in a zone from $z-\Delta z$ to $z + \Delta z$, where we have $kr \sim 1$.

Let's see a typical value of $\Delta z$. If the beam is approximately centered, x-coordinate is symmetrical so $\theta = \frac{\pi}{2}$. Consequently, in the expression of solution $A=F(r)P(cos\theta)Q(\phi)$, we see that all Legendre polynomials of even order are null and those of odd order are constants $< 1$. As $\phi \sim 0$ in figure \ref{nano_pos}, we have $Q(\phi) =e^{\pm im\phi} \sim 1 - im\phi $ and only the odd terms $h_{l}=h_{l}^{odd}=j_{l} + i y_{l}$ will survive. The evolution of real and imaginary terms shows that $j_{l} < j_{1}$ where $j_{1}(x)= -\frac{sin x}{x^{2}} - \frac{cos x}{x}$ is an oscillatory function with $y_{1}(0)=0$ and $y_{1} < 0.5$ for $x=kr \sim 2$, but also $y_{l}(x)$ is a growing suite for $x < \mbox{some }\lambda$ and have the same behavior as $j_{l}$ for big values of x. Typically if we restrict us to $m=3$, the frontier is $\Delta z = kz_{\delta}= 5$

Finally $h_{l}(kr) \sim i y_{l}(kr)$ (with always $x > \epsilon$ for avoiding divergence).

Now in the integrand will have also $\theta' = \frac{\pi}{2}$ if we suppose a TEM mode with vertical polarization. We also suppose that laser alignment is perfect, so $\phi' = 0$

In these conditions \begin{equation}A(\textbf{x}) \propto \sum_{l odd, m} a_{l} (1 - im\phi)*iy_{l}(kr)*F(z,r')\otimes[\int j_{l}(kr)Y_{lm}^{*}(\frac{\pi}{2}, 0)d^{3}\textbf{x'}]\end{equation}

    where the convolution by the presence function $F(z,r')$ is symbolically indicating which integral to compute depending on the locus of the sources relatively to the particle inside the three domains.

    Simplifying for $Y_{lm}^{*}$,

    \begin{equation}A(\textbf{x}) \propto  \sum_{l\mbox{ }odd, m} a_{l} (1 - im\phi)*iy_{l}(kr)*\int j_{l}(kr) d^{3}\textbf{x'}\end{equation}

    The $d^{3}\textbf{x'}$ domain may be divided in two, inside and outside the zone $[-\Delta, \Delta]$, for summing on z, neglecting the immediate entrance and sortance of the electron. Indeed, in that case,  the associated phenomena stays identical whatever the angle $\phi$.

    There is no necessity to calculate here, the fields components, because we can \textit{a posteriori} directly inject the vector potential in the hamiltonian of interaction, ie in a scheme in which the particle is present.

\textbf{Note} : until now, we rely on the evanescent wave acceleration, to imagine laser acceleration in nano structures. Although it is a practical basis for our system, it is not today demonstrated to be the \textbf{unique} way of field acceleration violating Lawson law, other geometries and coupling could to be evaluated, for instance a ribbon 

    \subsection{Electron optics}   

    Inside a RF standard gun for instance, with relativistic output, the longitudinal accelerating electric field is predominant against transverse one. Moreover one can show that Lorentz transformation results in reinforcement of $\frac{E_{l}}{E_{t}}$ in the particle reference coordinates. Conversely for low energy sources at 10keV, we are not benefiting from that situation. How to alleviate it ?

\begin{figure}[h]
\center
	\includegraphics[scale=0.4]{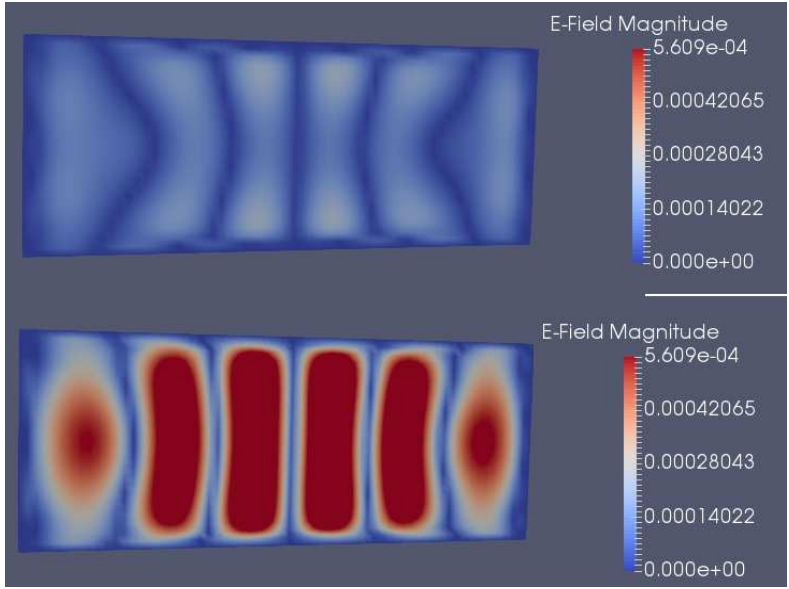}
\caption{Efield transient tip Response to laser gaussian 100fs pulse, opposite plane, times 95 (up) and 139fs}
\label{screenshot}
\end{figure}

    Firstly, at very near near distances from the emitting tips, the electric field, specially the longitudinal one, is very intense. For information only, the figure \ref{screenshot} represents two screenshots of the time-space evolution of electric field magnitude diffracted by an individual tip ($h=2\mu m, apex=10nm$) at $t=95fs$ and $t=139fs$ for a gaussian laser pulse of 100fs width. The simulation was driven under OpenEMS and animation results with Paraview. The observer distance is $2\mu m$.

We evaluate the field amplification $\beta \sim 11$. The gap cathode-anode is 3mm. The peak macroscopic electric field is 3.3M V /m with tip, E Schottky field is -without tip- E gap = 36M V /m. The \og{}amplified\fg{} laser field is evaluated to 1.28GV/m without tip amplification, and to 14.1GV/m with tip.

That field behavior, evaluated by electromagnetic FDTD tool \cite{openEMS}, brings us to distinguish the near field zone where the bunch experiences a strong accelerating longitudinal forces, from the remaining space inside our -miniature-gun, where the acceleration is not sufficient to preserve the bunch cohesion. The typical cathode-anode (iris) distance of our the gun being 3 mm in first draft, large in front of near field one, there is then a critical distance generally lower than $1 \mu m$ beyond that it becomes necessary to focus the beam, before transporting it through anode iris and presenting it at DLA aperture.

\begin{figure}
    \begin{minipage}[t]{7cm}
        \centering
        \includegraphics[width=6cm]{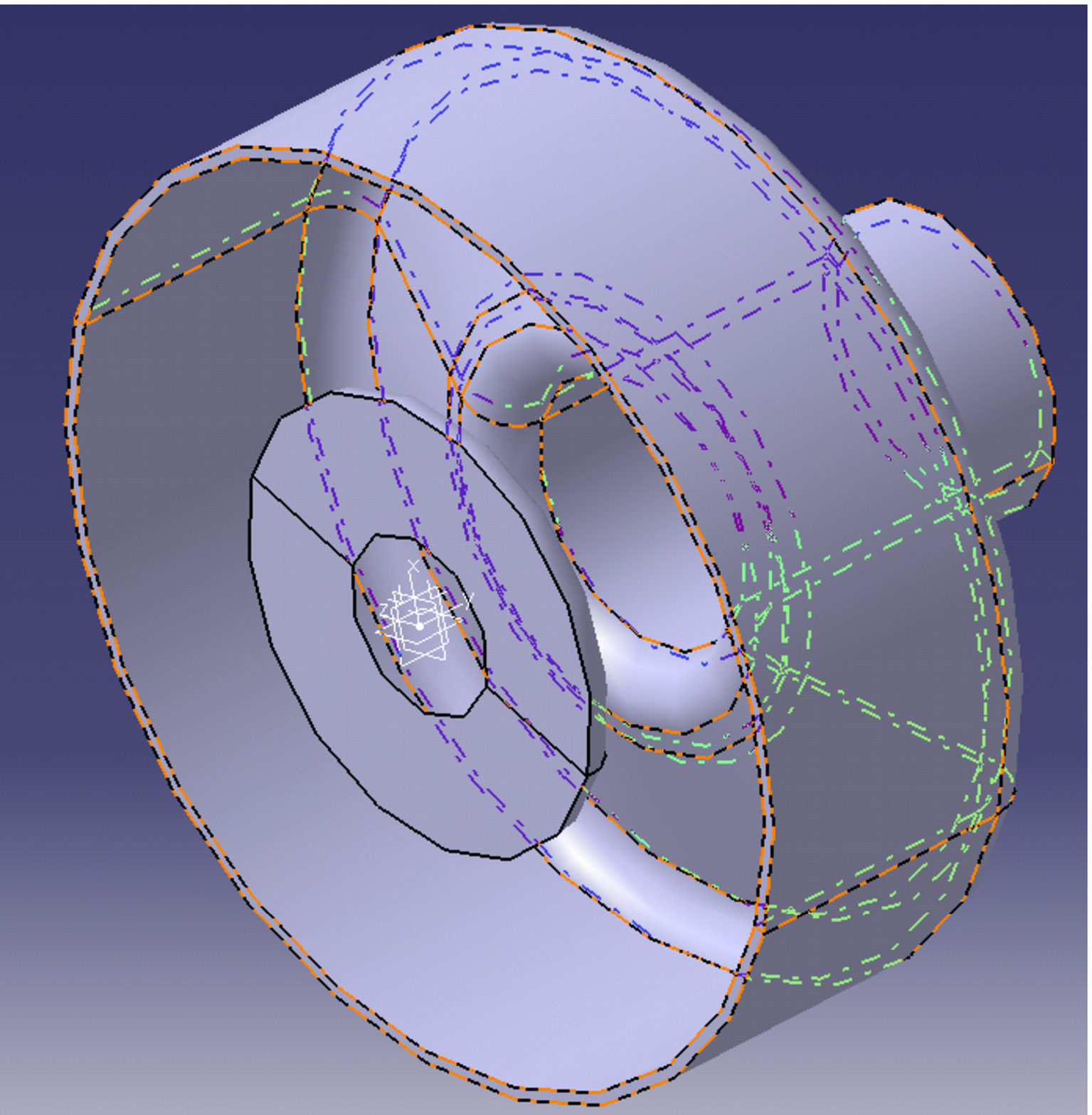}
        \caption{Coaxial gun version}
        \label{canon_v12}
    \end{minipage}
    \begin{minipage}[t]{7cm}
        \centering
        \includegraphics[width=6cm]{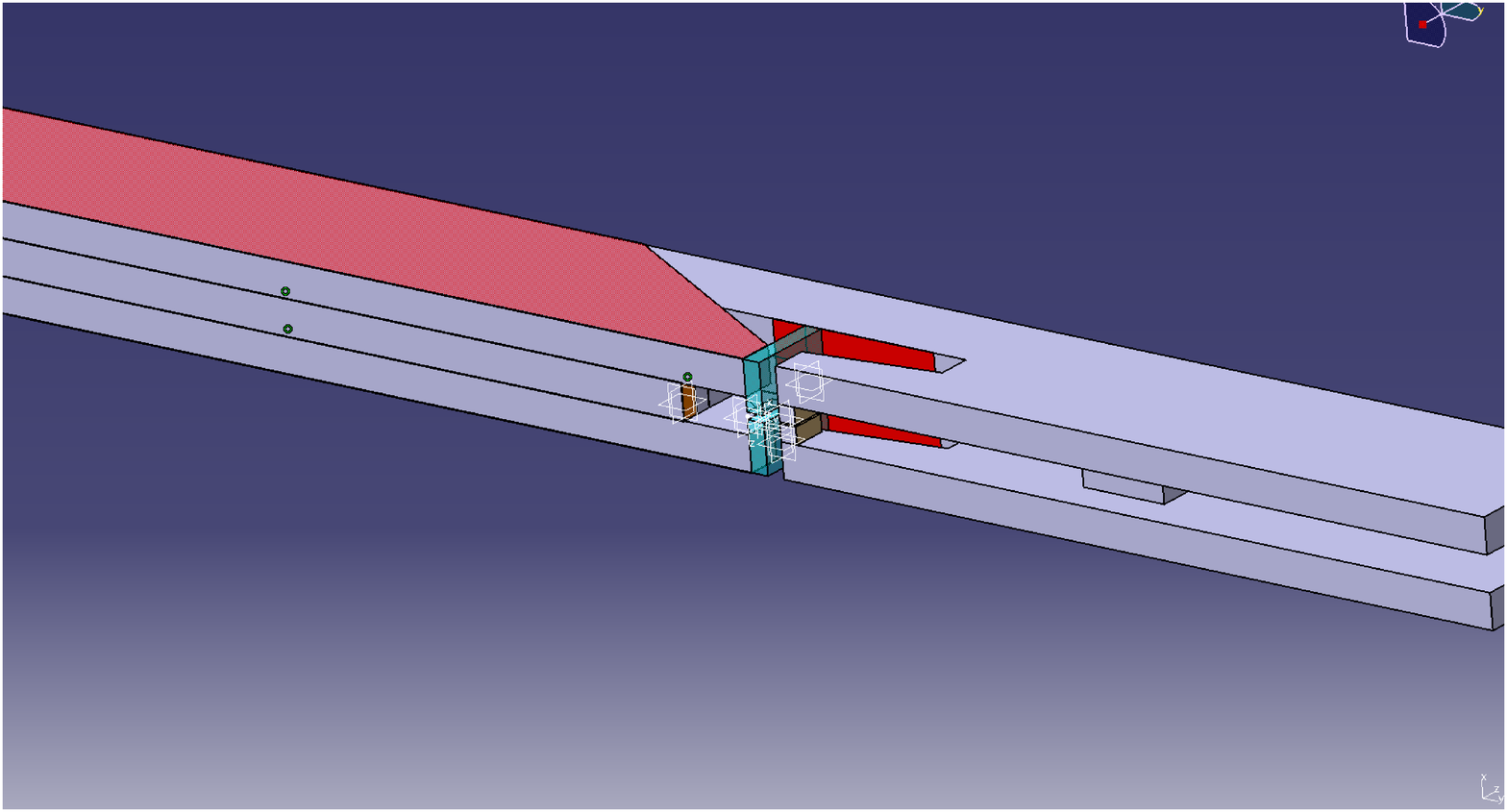}
        \caption{Integrated version}
        \label{canon1}
    \end{minipage}
\end{figure}

    A standard focusing optic can't be easily realized inside 3mm typical cubic volume with miniature focusing apparatus, and more generally will be a prototype in itself, inside the dimensions of our setup. Moreover, the auto induced stray fields at decimeter extensions of a magnetic or electrostatic lens, could be a very crippling problem. Respecting our \og{}game rule\fg{}, ie \textit{only} low cost laser, nanostructures and dielectric arrays,\footnote{although some millimetric size realizations has been proposed, nevertheless, these setups necessitates focusing electrodes, so don't obey to our game rule} we propose to insert a DLA module inside the gun, at the critical distance of FEA. That proposal scheme is represented with two variants, with and without focusing optics. In coaxial version, figure \ref{canon_v12} we have designed a first draft of cylindrical profile, which should have some auto-focusing capabilities. We followed some guidelines, not detailed here but it is to refine further thanks to softwares. In integrated version, figure \ref{canon1} the very short distance between FEA and DLA avoids the focusing need. Moreover, the strip-line disposition allows to choose a proper input impedance, using the synthesis with Hammerstad equations.

We may ask wether it is convenient to immerse in dielectric transparent to 10keV electrons and to laser wavelengths, the zone between FEA and DLA, and the inside of DLA. From one side, fabrication may be easier, and focusing -better or not- is questionable. \footnote{In such an entirely integrated device, looking like a semi-conductor, it could be possible to work without vacuum environment} From other side, the occurrence of collisions of the bunch by centers inside the crystal, essentially inside a length of roughly 1 mm, has to be closely studied in order to determine such a feasibility and the induced losses.

\textbf{*} : electrical/electromagnetic behavior of that 1D device remains to explore. We know that depending on laser phase, deflecting, focusing or accelerating forces are impinging on the bunch. Moreover, the linear FEA necessitate asymmetric lens so we must apply specific electron optic.

\textbf{*} : as precised in \cite{Born}, (appendix II in 7th edition) , the wave propagation of electron bunch is not driven by a simple eikonal, the direction of trajectories is not orthogonal to equiphase surfaces, but is rather parallel to electron momentum (which is not necessarily orthogonal to these surfaces). So we see up to now that we shall have to represent the bunch dynamics at low energies, by a Lagrangian approach. Appreciation of Lagrangian or Hamiltonian choice -and even hamiltonian dilemma due to representation of excited states- is left to further discussion. Moreover, the landscape is obscured by charge space and many-bodies interaction inside bunch. All these aspects should be integrated in a single model, also left for next work.   

    \subsection{DLA first stage}

    \subsubsection{THz cells}

    In Thz acceleration, the laser frequency is divided to THz range and interact with the bunch inside a cell with mm dimensions. In such system, a 100 fs laser pulse, for instance, is rectified such the electron bunch \og{}sees\fg{} an approximate laser accelerating amplitude during its travel inside the cell. It has a great advantage of a strong acceleration distance and the energy gain may be very high for a 10keV bunch, imagining final energies near relativistic range with a single cell. THz cells benefit from existing experiments. 

    However that process presents some drawbacks :

    \begin{enumerate}
    \item Division of laser frequency inside active crystals, so limitation of incident power, and supplementary cost,
    \item Large dimensions of that first stage, not very coherent with in-chip philosophy
    \item Necessity of focusing between photocathode and THz cell, because the dimensions of that cell don't allow to realize an integrated device
    \end{enumerate}

Regarding the optical rectification, the necessary energy input for a standard ZnTe crystal is in the $1\mu J$ range for a consistent second order non-linear yield, far from the nJ energy of the source we intend to use, and far also from the Yb fiber lasers planned for the in-chip accelerator. To get that conversion, we might use an ND-Yag amplifier but its PrF is limited to 10Hz.
    
In the following, we deal with non-THz cells. However, THz techniques may be useful and are considered inside \ref{Measurements} section.
    
    \subsubsection{Traveling wave and direct acceleration in near infrared laser range}

    \paragraph{Analysis}
    
    In a first step, standard double grating is planned for our experiment, and its geometry is naturally adapted to the proposed 1D flat FEA. In all the following, we shall restrict us to traveling devices, so THz cells are not studied here, they constitute a very different landscape because we cannot consider the material tailoring with propagation. Instead focusing us to propagation we must study the cavity aspect and these structure are fundamentally different from that fact.

    However, let's discuss here more generally of freedom degrees for structure fabrication, geometry, material choice, \ldots of an \og{}ideal\fg{} sub-relativistic accelerating stage. Before dealing with it, we point out some interesting topics and research directions.

    The reference \cite{Hughes} about AVM method, illustrate in another point of view, the mutual interaction between the beam and the total induced field inside accelerating structure. Although innovating computation method, it reminds us the well-known problem of beam loading for example and has to be kept in memory in any design; furthermore, it must start from a deliberately chosen initial design. In other side, much work has been done for optical guides, specifically with ribbons, not directly transposable for the main e- channel, but at less for light couplers \cite{Drachenberg}.

    So in any design, the two quantities $F_{a}$ as in \cite{Hughes}, and normalized emittance $\epsilon$ must be specified.\footnote{without degrading generality, for fixing ideas, let's choose for instance, $F_{a} = \frac{G}{\frac{1GV}{m}} \sim 0.5$, where G is acceleration gradient and E maximum field, and $\epsilon \sim 1nrad.m$}

      Now, we dispose of a spatial and temporal profile of entrance bunch, transported through a short critical distance, knowing it is generated by a 1D flat FEA array, we have a set of dielectric volumes, of  arbitrary indexes. \footnote{we don't prohibit meta-materials} There is no reason to suppose that sectional XY bunch profile \footnote{X is chosen for 1D axis of FEA, then Y is vertical axis} is uniform in X, and of course in Y. Nevertheless, the held symmetries are relative to axis OX and OY, so our device is not 2D equivalent but a volume, and may however keep these quadrant symmetries. Its length is determined by constraint on distortion, then output emittance. It may be evaluated by independent simulations, so the main problem is the determination of sectional $S(z)$ structure of our DLA.\footnote{Rigourously we must talk about $S(z,t)$}

      Also, the suite $S(i)$ takes in account the energy variation of the bunch, like in the case of chirped gratings. The entrance section $S(0)$ is not a rectangle but must be adapted to entrance profile of the bunch. Which profile have we to choose, knowing that the bunch profile XY is not constant from head to tail ? There are two strategies :

      \begin{enumerate}
      \item the entrance $S(0)$ is vacuum and is sized for maximum section of the bunch,
      \item $S(0)$ is partially in vacuum or not at all, for its central part, and its near boundary regions are made in dielectric material, offering a multi-layer index $n(r), r>r_{0}$
      \end{enumerate}

      The first strategy is widely used, for instance for the double grating. The second one is somewhat new, although already used in resonant devices. The beam is characterized by a sectional distribution of charges and momenta.\footnote{we admit to restrict our generality to cartesian coordinates. It is possible that coordinates adapted to equiphase or momentum gradient surfaces of the e- beam be more pertinent}

      The other important freedom degree is linked to laser injection and polarization. Escaping the technology influence, we define a electric state of the laser field at the frontier of DLA, independent of realization of couplers. We make the hypothesis that notable component of laser field must be present at all steps of accelerating path, the simplest case being an uniform injection in amplitude.\footnote{generalization with non constant field is another difficult step}. Restricting the generality, we study the possibilities of transverse and longitudinal laser injections, relatively to electron mean axis beam. laser e- interaction may be indirect, by modes response of the material.

      For short, the limitations and rough hypothesis of our study are the following :

      \begin{enumerate}
      \item laser injection is transverse or longitudinal
      \item the DLA device has  Ox and Oy quadrantal symmetries,
      \item for beam propagation, we are allowed to cut it in slices of volumes $S(z)\delta z$, each one having a profile n(x,y,i), n being the complex index (including meta-materials), and $z=i \delta z$
       \item for laser injection,\footnote{even that simplification raises problem because the incident equivalent wave front of electron bunch is not a plane, so cutting in cartesian slices is not the best way to deal with propagation. As a justification, we can say that each slice thickness may be chosen enough tiny to push in second order, the \og{}aberrations\fg{} of the electronic wave front, and to dispersion inside optical cycle.}
      \item figure of merit are final parameters at DLA output, given emittance and $F_{a}$
      \end{enumerate}
        
\begin{figure}[h]
\center
	\includegraphics[scale=0.4]{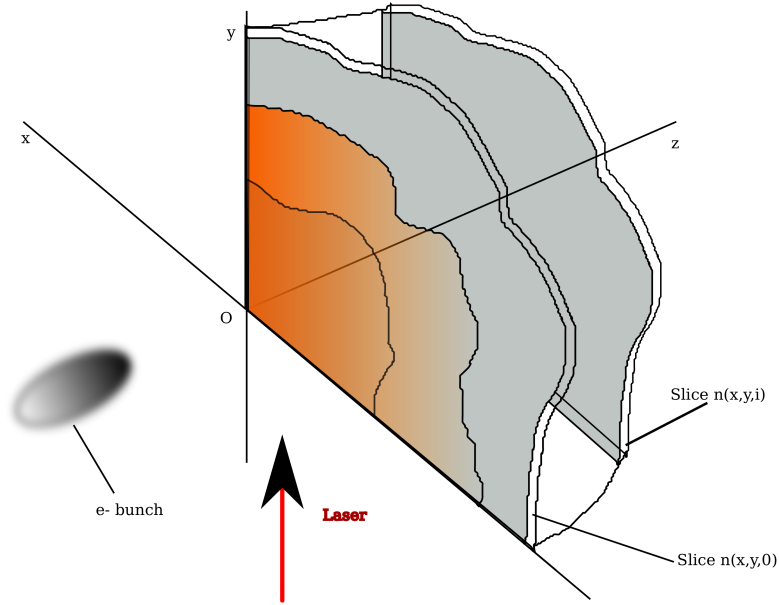}
\caption{General case of Dielectric Laser Acceleration}
\label{idealdla}
\end{figure}

      The most general scheme, under our precedent constraints, is suggested by the figure \ref{idealdla}

\textbf{*} As in standard interactions models, we have to describe the laser field, the material response to bunch transit, taking account the $n(x,y,i)$ suite, and define a cross mechanism between them. 
      
      \paragraph{Material choice}

      In fact, we have not an infinite freedom in materials choice, but it should ask for a future comparative study. Dielectric acceleration lays on high field strength and high flux dielectrics, silicon, AlTiO3,...In order to gain supplementary performances and use complex indexes, we may think of meta-material and photonic combinations with these.

      Regarding meta-materials \cite{Lanneberre}, the ceramics like TiO2 for instance present negative permitivity, low permeability and consequently a negative index, giving them properties of magnetic mirrors in relatively narrow bands near 500GHz. We must keep in memory that an optical pulse of carrier frequency 400THz and of width, say 100 fs, has an equivalent frequency band of $0.35 \frac{1}{t}=3.5 THz$, far above the range of these meta-materials, so in the reference it is suggested for high frequencies, to use rather polaritonic materials. Indeed, their frequency range may fit the optical laser spectrum, but there is yet no proof they should be adequate for high fluxes and high electron energies, as most publications refer to micro cavities with some eV band gaps.

    \subsection{Measurements}\label{Measurements}

    Asked bunches characterizations are numerous, but the minimal set is :

    \begin{enumerate}
    \item mean current (total charge/second)
    \item bunch duration or length
    \item bunch position and spatial lateral dimensions of the beam,
    \item energy spectrum
    \end{enumerate}

    We emphasize that measurements are not integrated part of our \og{}game rule\fg{}, which is devoted to the electron source, so at the outside of vacuum cell, we obviously shall find measurement devices. It seems that effort to precisely define diagnostics for that setup are in nearly same amount that those defining the source in itself. A reason why is the inverse conversion from fast bunches to temporal electrical signals. Ideas are either transposition of standard techniques for long bunches, or indirect methods by photonics. We can extrapolate one of them, \cite{Kozak}, using a decelerating grating, instead of deflecting one, the decelerating grating plays the role of energy high pass filter. That trick permit measurement simplification, all is done in the axis, and the same laser oscillator allows accelerating and decelerating phases. However it is necessary to implement two DLA on the same wafer.

\subsubsection{Charge of a bunch}
    
    Mean current will be known by transposition of standard techniques like Faraday cup. However, in our very low charges context, we guess it may present several issues. We have analyzed the reference \cite{Harasimowicz} which gives interesting design elements, even applied to ions. Indeed, we retain the cup geometry and the necessity to do some simulations; we shall have also to evaluate if cooling is or not necessary. Following the reference, we evaluate roughly the total charge and thermal eventualities. If we want to capture the entire beam, without extra focusing, the entrance slit of the cup must stay at very tiny distance from DLA output. By analogy with the temporal dispersion, we know that roughly, a 10keV femtosecond electron beam without manipulation, may be enlarged fastly to 1 ps. The spreading distance, accounting for 0.2 c speed, will be $d=1ps * 0.2 c = 60\mu m$. It is of course unrealistic, unless to \og{}envelop\fg{} the output by the cup. In other side, Faraday cup is claimed to be able to measure pA currents. In our case, the charge estimated at the entrance of DLA is to be 34mA, we calculated that a 1D Array of 1000 tips, each one delivering $34 \mu A$, is convenient. The charge collected by a 100 femtosecond pulse has been also evaluated to 0.27 fC, so the mean current by pulse will be $<i>_{pulse}=\frac{0.27 10^{-15}}{100 10^{-15}}=2.7 mA$. Obviously, that current is by far, above lowest limit of that device, but it is to be verified that the low charge itself, is not a limiting factor, because of possible noisy -fast- charging effects. Of course, with MHz PrF, the situation becomes comfortable.

    That device, although of reasonable dimensions, say 5cm, kills the beam, so either it must be removable or micro positioned in order to move it away from the beam inside the vacuum cell; alternatively we may realize a set of integrated DLA(s) with and without Faraday cups directly printed on chip. For not integrated solution, a - low-cost - micro positioner appears to be useful.
    
\begin{figure}
    \begin{minipage}[t]{7cm}
        \centering
        \includegraphics[width=6cm]{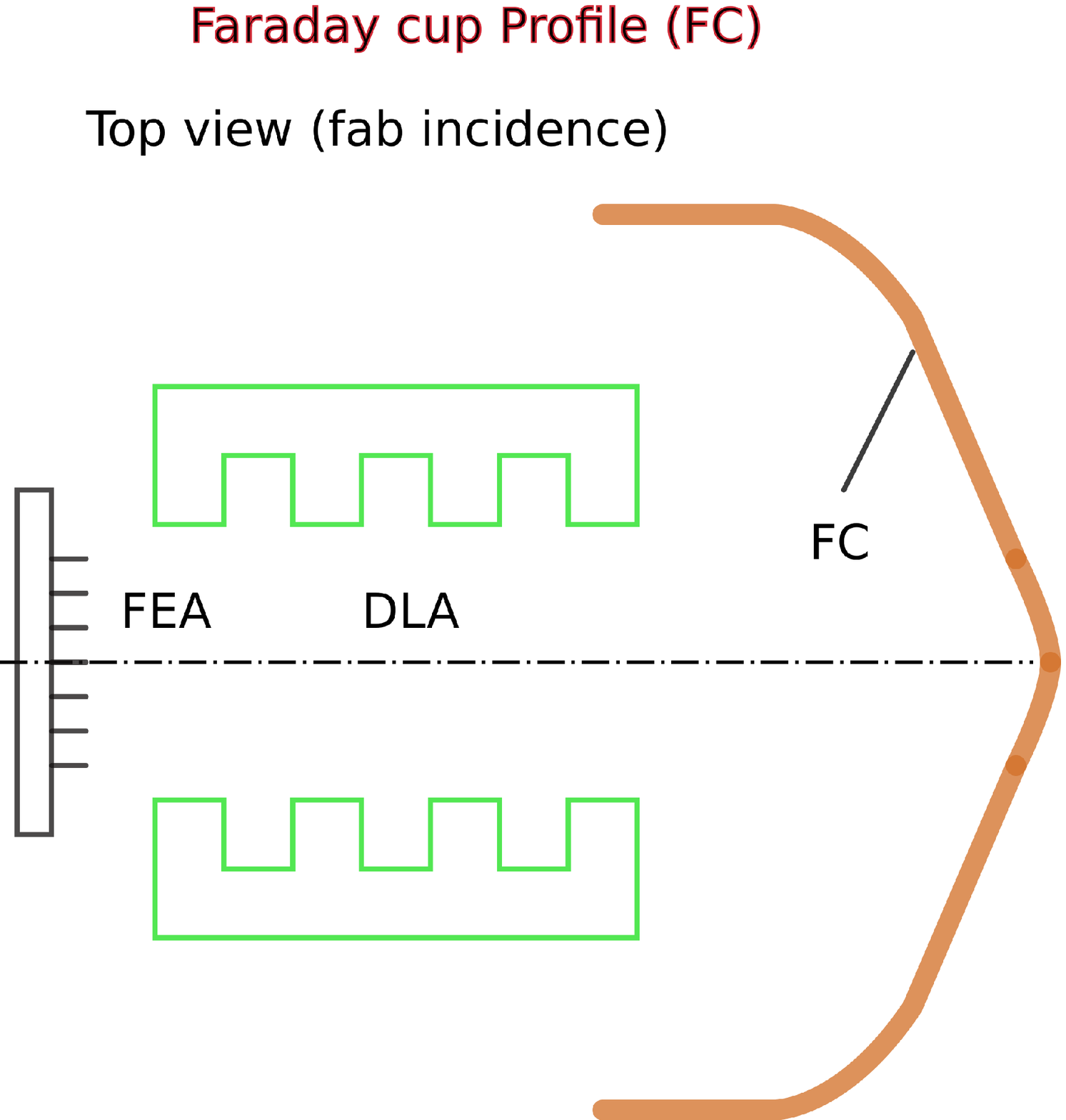}
        \caption{miniature Faraday cup}
        \label{cup}
    \end{minipage}
    \begin{minipage}[t]{7cm}
        \centering
        \includegraphics[width=6cm]{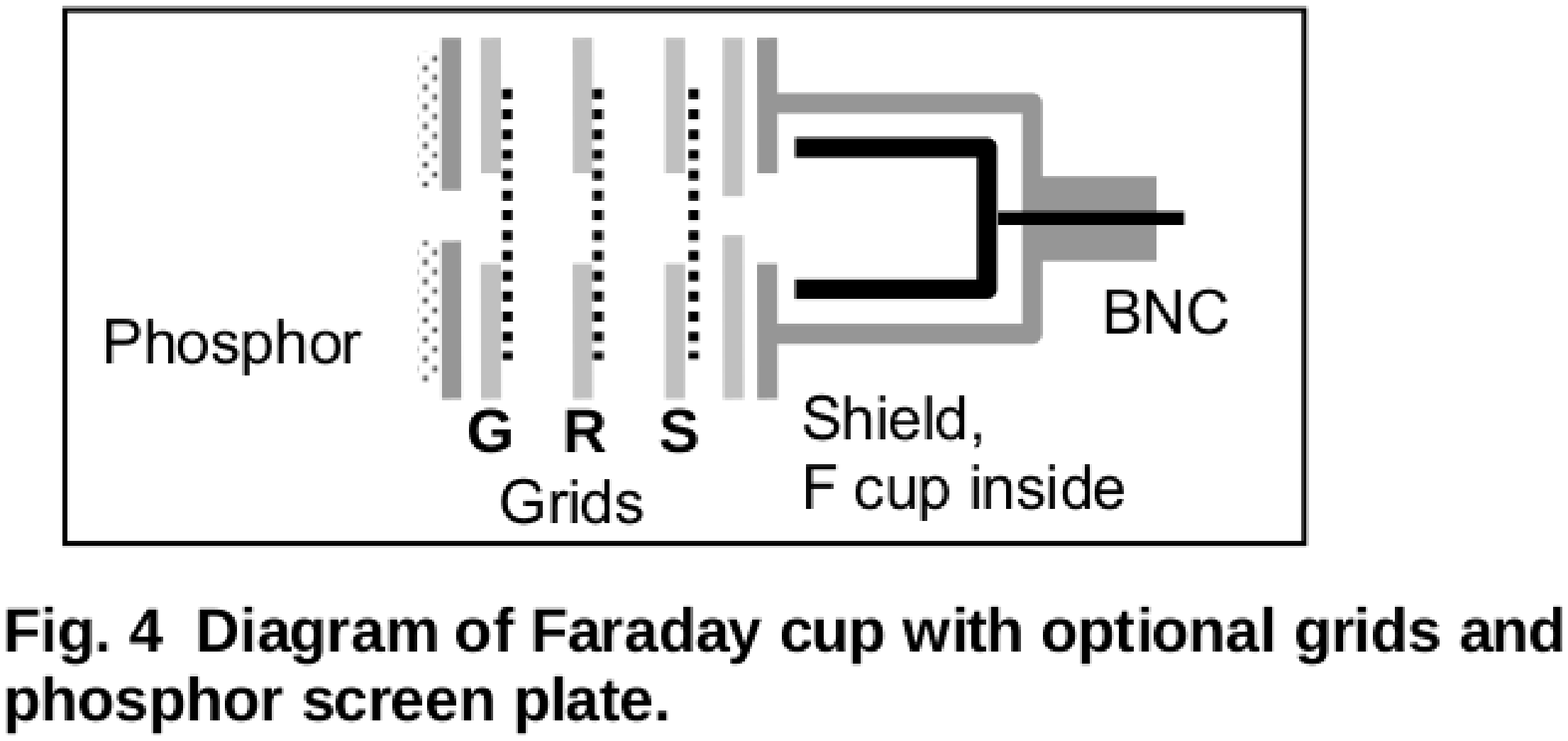}
       \caption{Standard device (from Kindall)}
        \label{Kindall}
    \end{minipage}
\end{figure}

Among the issues raised by the implementation of ultra low currents, the reference points out the noise induced in the measurement cables, the vibrations of vacuum pump and the residual components from vacuum to air feed through(s). Exploiting the schematics of the reference, and those from other standard realizations of CERN, we propose a setup by the figure \ref{cup}.

 Regarding noise sources, we thrust in vacuum team to choose the adapted vacuum pumps and materials, and feed through. We concentrate on electronic components and we suggest using optical fiber for signal line. its powering may be brought by the polarizing ring, and we found performing fedoras for fiber optics up to $10^{-9} Torr$. The signal is digitized inside the apparatus and converted to optical one. The stopping material could be low thickness graphite - stopping distance is quasi null at 10 keV - partially covering the DLA output, because Coulomb scattering is intense at this point.\footnote{without Faraday cup, a photonic focusing lens is then necessary for the majority of other measurement} Another solution is to integrate Faraday cup in-chip, and replace sample by loadlock, one of them having integrated Faraday cup, others not.

We notice that the Faraday cup may be used for other measurement like energy spectrum, modulating the repelling voltage grid for secondary electrons. Nevertheless that apparatus doesn't allow recording sub picosecond fast signal. A phosphor screen could be inserted for measurement of transverse dimensions of the beam. The figure \ref{Kindall} recalls the standard product.

\subsubsection{Bunch length}

Thz techniques have already prove their efficiency for picosecond bunches \cite{Wilke}, we propose to extend them to femtosecond measurements. Among the main useful parameters are the first phonon resonance conditioning the bandwidth, here 3.5 THz for ZnTe- then enough for 100fs measurements as we saw previously. Other crystals like GaAs may present better bandwidth, and it was showed that 50fs lengths are accessible by these measures. For ultra-short bunches from as to 10fs range, \cite{Reckenthaeler} propose a streaking camera based on Auger effect. It seems that Titanium target is well fitted to that measurement, but of course that technique is beam killing.

\subsubsection{bunch position and lateral dimensions}

Photonic techniques seems to be adapted to these measurements, if we exploit the proposal of \cite{Soong}. As showed in that reference, the signal is essentially sensible to one dimension only, so Implementation may be in X and Y if we use two gratings mutually rotated of 90 degrees, first one for X and second one for Y. However, the processed signal is devoted to X,Y position and may also take account of a point like particle, then for example a \og{}center\fg{} of a bunch. If the bunch is extended in size, complexity may rise in processing. However that situation is encountered also in standard electromagnetic BPM, so solutions are to be found.

That section is deliberately incomplete. Diagnostics is a full and difficult domain in itself and is to be completed before practical steps.

\subsection{Summary and discussion on possible and preferred FEA and DLA configurations}

\begin{figure}
    \begin{minipage}[t]{14cm}
        \centering
        \includegraphics[width=12cm]{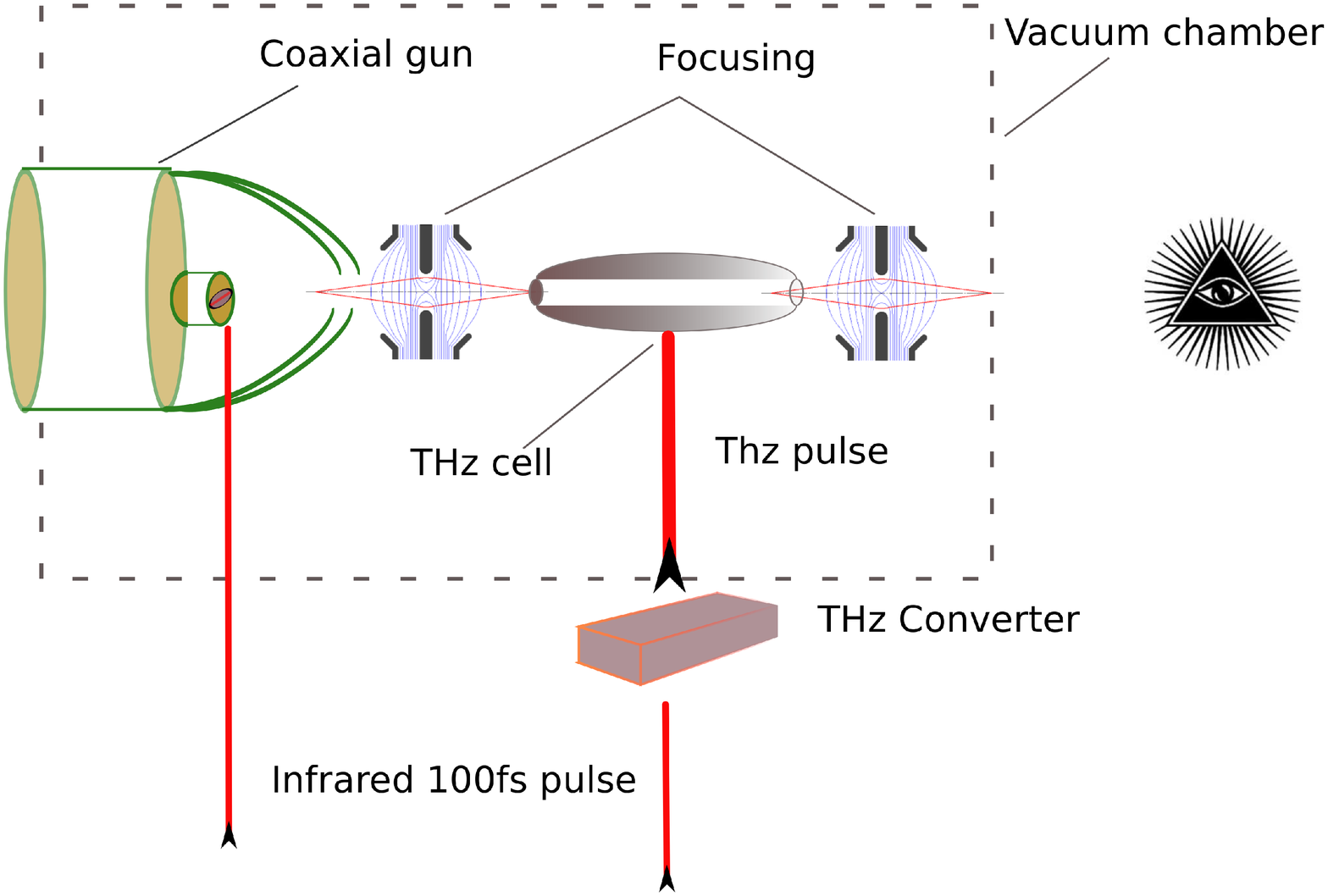}
        \caption{Setup with THz cell}
        \label{solution_a}
    \end{minipage}
    \begin{minipage}[t]{14cm}
        \centering
        \includegraphics[width=12cm]{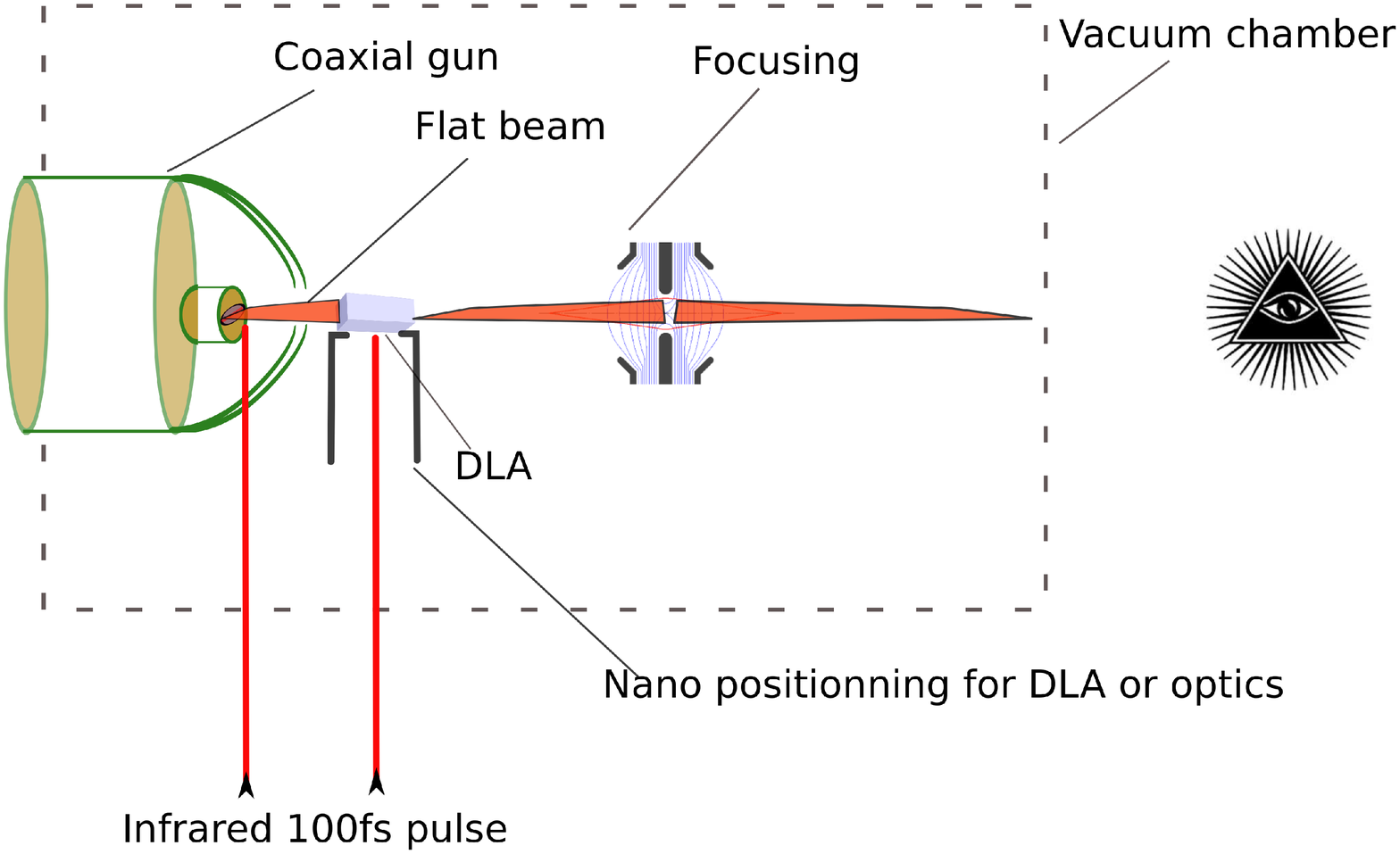}
        \caption{Coaxial setup with DLA}
        \label{solution_b}
    \end{minipage}
    \begin{minipage}[t]{14cm}
        \centering
        \includegraphics[width=12cm]{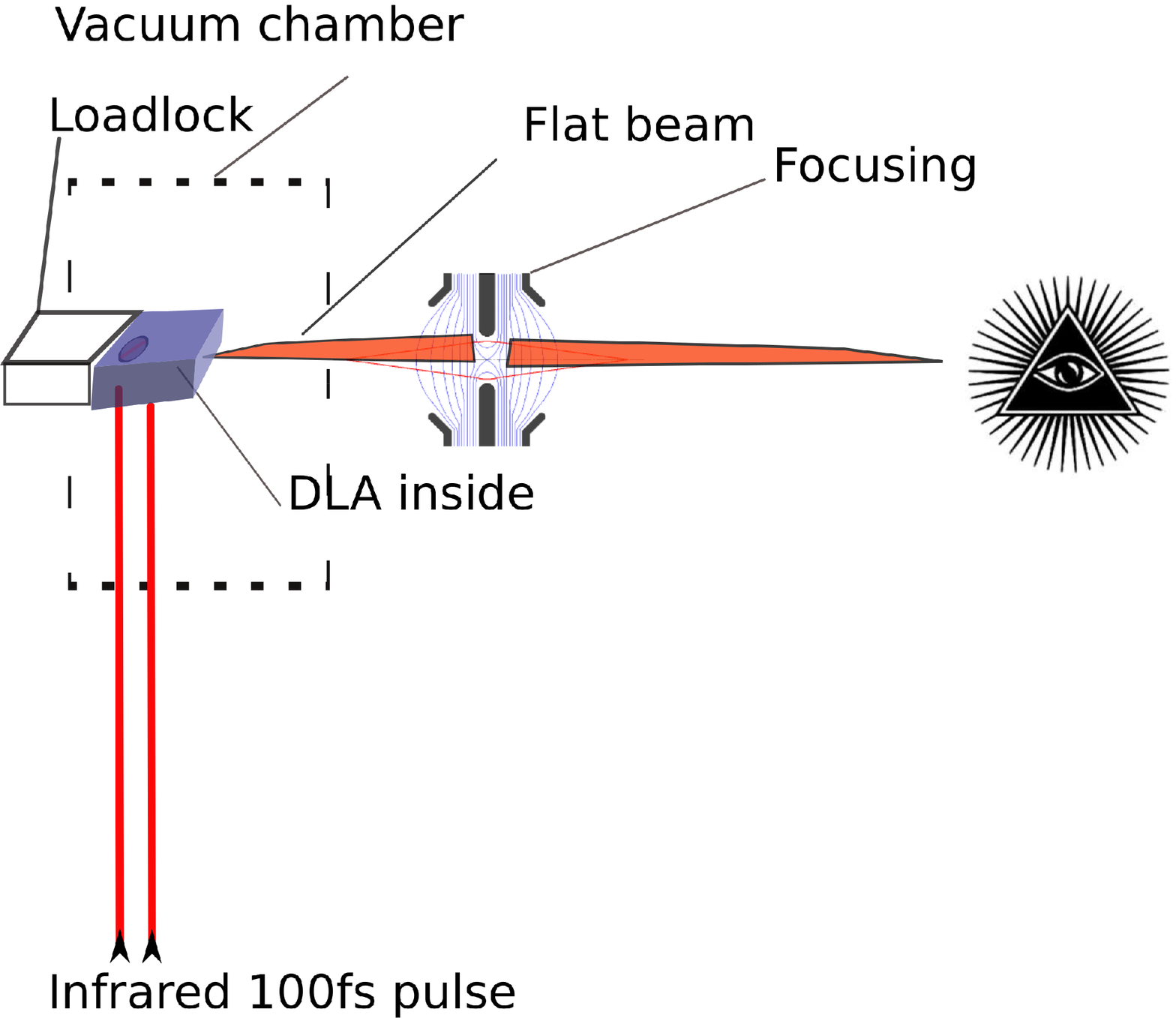}
        \caption{Integrated setup in strip line configuration}
        \label{solution_c}
    \end{minipage}
\end{figure}

The precedent sections enlightened several possibilities of integration of the couple FEA/DLA, which are summarized on figures \ref{solution_a}, \ref{solution_b} and \ref{solution_c}.
In all layouts, the \og{}eye\fg{} encloses all measurements apparatus, and is outside of vacuum chamber, if any.

Obviously, Thz cell (solution a) although ideally efficient in energy, leads to cumbersome setup, difficult to tune, and generating a growing emittance due to several drift and focusing spaces, at least for femtosecond bunches. It is also necessary to convert the laser frequency to THz, with enough flux for the DLA.

The solution b was our first idea, because our electrical source was planned to connect to coaxial termination. We have seen that nano-positioning of the DLA relatively to FEA may be extremely difficult. Even with nano-positioning of optics, it rather leads to a demonstrator, and can't become a realistic -perhaps industrial- solution.

A more refined stage consists logically to the integration by nano-fabrication of FEA and DLA. Nevertheless, the coaxial setup renders uneasy the laser injection, is mechanically complex (even possible, we have studied it), and the anode distance can't be reduced to micronic range, so divergence of the beam must probably be constrained before next DLA stage.

Finally, our contribution, solution c, emphasizes the interest to develop the last solution d, with entirely integrated chip, in a PCB like configuration. The PCB like layout with a buried symmetric strip-line for injection, is compatible with ultra fast sub-nanosecond electrical HV pulses, or RF amplifier output, and we are able to tune the cathode impedance. It seems at first glance that it impedance is easily tailored with strip line, easier than with coaxial cables. However, with the geometries that we dispose, the calculated impedance are low, so either we rise then by some trick, or we deliberately design a short circuit stub. In that last case of course, numerous issues appear, even if we master nano-structure fabrications.

\newpage
\section{Conclusion}\label{conclusion}

Our contribution is a proposal for a future sub-relativistic 10keV range electron source, compatible with Laser acceleration module made in nano-structure, Dielectric Laser Acceleration (DLA). We have described and justified an experimental setup, more advanced than initial demonstrators given in literature, our proposal goes directly to validation of the system.

Its components are based on a Field Effect Array (FEA) flat beam photo-cathode, functioning in field emission pulsed-DC mode with RF or ultra-fast HV generator, associated in integrated fabrication with a first stage DLA module. We estimate that fabrication with very short distances between FEA and DLA, brought naturally by integrated fabrication, will avoid the need of additional optics. Overall system should hold inside a decimeter size chamber, including new measurements of beam based on photonics. A future improvement should be to integrate the laser couplers to DLA.

In order to reach demanding performances, it is necessary to evaluate new candidates as Carbon Nano Tubes (CNT) for the FEA configuration, but their assembling must be done in link with solid state and surface physics experts.

We have left out number of theoretical and simulation issues for other works, to list them, not exhaustively :

\begin{enumerate}
\item Initial momentum and energy distribution of electron at the immediate interface, resulting from cathode physics and estimated by \textit{abinitio} models and simulations,
\item Presence of surface states and influence on ulterior bunch dynamics,\footnote{intuitively, the surface state presence/predominance is associated with very low energy vacuum emitted electron, those which stay in meta-stable state in the \og{}turning points\fg{} of the potential well. It has to be determined if they have significant decay time in a non equilibrium scheme like ultrafast photo-emission} 
\item estimation of thermal stresses on cathode under high PrF and emitted charges,
\item electron dynamics in low charge regime, with eventual many-body interactions, and setting of a general model,
\item optics evaluation with flat beam and nano-structures,\ldots
\end{enumerate}

For a general model of emittance dilution including many-bodies if necessary, certainly some numerical developments do exist, it should be interesting to compare an analytical Lagrangian approach with them. Among next contributions it should be useful to analyze the precise case of sub-femtoC bunches in term of several characteristic length of that pseudo-plasma, and deduce if eventually an hydrodynamic model is -or not- adapted to these dynamic evolution.

We also point that much stress could be reported on performances of clean room processes. A draft proposal program for fabrication has been transmitted to our specialized clean room platform (Institute of Fundamental Electronic, IEF) at Orsay. Outside expertise is of course very welcome in that complex topic.\newline\newline

Thanks to : the representatives of the society Physik Instrumente (PI) who gave me elements of nano positionning ; Sandry Wallon at CNRS/Lal/Depacc for his generous help and knowledge in that topic ; Moana Pittman at CNRS/Lal/Laserix for preliminary experimental discussions around Laser injection.

\newpage

\bibliographystyle{unsrt}
\bibliography{keV_DLA_manip}

\end{document}